\documentstyle[aps,prl,manuscript]{revtex}
\begin{document}
\draft
\newcommand{\vektor}[1]{\mbox{\boldmath $#1$}}
\title{Finite temperature bosonization}
\author{Garry Bowen and Mikl\'{o}s Gul\'{a}csi}
\address{
Department of Theoretical Physics, 
Institute of Advanced Studies \\
The Australian National University, 
Canberra, ACT 0200, Australia}
\maketitle
\begin{abstract}

Finite temperature properties of a non-Fermi liquid system is 
one of the most challenging probelms in current understanding
of strongly correlated electron systems. The paradigmatic arena 
for studying non-Fermi liquids is in one dimension, where the 
concept of a Luttinger liquid has arisen.
The existence of a critical point at zero temperature in 
one dimensional systems, and the fact that experiments are 
all undertaken at finite temperature, implies a need for 
these one dimensional systems to be examined at finite 
temperature. Accordingly, we extended 
the well-known bosonization method of one dimensional 
electron systems to finite temperatures. 
We have used this new bosonization method to calculate 
finite temperature asymptotic correlation functions for linear 
fermions, the Tomonaga-Luttinger model, and the Hubbard model.  

\end{abstract}

\newpage 

\section{Introduction}
\label{intro}

The many-body problem in condensed matter physics generally 
relies on the machinery of perturbative quantum field theory 
to obtain solutions to model systems.  
Solutions to the field theories are obtained in terms of 
{\sl Green's functions}, or equivalently, {\sl correlation 
functions}.  The Kubo formulae relate the retarded Green's 
functions of a model to the linear response of the system 
due to external fields.  The experimental responses of a 
real system may then be compared to the solutions for the model.

Experiments on many-body electron systems normally measure 
properties at energy scales small compared to the Fermi energy.  
This means that only a few degrees of freedom in the system are 
excited, and only the low energy sector of a model need be compared 
against experiment. 

The low energy, long distance physics also determines any long 
range order and cooperative phenomena present in a system.  
Divergences in certain correlation functions are indicative 
of the presence of ordering phase transitions in the system, 
such as ferromagnetic transitions or the Cooper pairing 
superconductivity phase transition.
The field theory methodology has been extremely successful, 
with the properties of a number of simple metals accurately 
modeled by using Landau's Fermi liquid theory.
However, many materials have emerged that have non-Fermi 
liquid properties.  

The paradigmatic arena for the study of non-Fermi liquid behaviour 
is in one dimensional many body systems. Simple dimensional 
arguments show that whilst finite momentum transfer 
interactions become irrelevant at low energies in higher 
dimensions leading to Fermi liquid behaviour, in one dimension the 
finite momentum transfer interactions are marginally relevant at 
all energies.

Normal perturbative methods of solution can no longer be 
justified in one dimension.  This is because the non-interacting 
one dimensional fermion gas is unstable against the switching 
on of interactions.  This is a form of the orthogonality 
catastrophe, where the interacting ground state is orthogonal 
to the non-interacting ground state. For this reason, 
non-perturbative methods of solution are now being examined.

Ground state properties of one-dimensional Fermi liquids 
are very special in that they retain a Fermi surface (if 
defined as the set of points where the momentum 
distribution or its derivatives have singularities) 
enclosing the same $k$-space volume as that of free 
fermions, in agreement with Luttinger's theorem. However,
there are no fermionic quasi-particles, and their elementary 
excitations are instead bosonic collective charge and spin 
fluctuations dispersing with different velocities. An
incoming electron decays into such charge and spin excitations
which then spatially separate with time (charge-spin separation).

Spin and charge separation is seen in all one-dimensional 
electron systems and it seems that it is not related to 
other characteristics of the one-dimensional phenomena. The
reason for this is simple: spin and charge separation is
due to scattering at the same Fermi point. Accordingly, it 
will not give any contributions to effects which are due 
to scattering between two different Fermi points, e.g., 
the renormalized correlation function exponents. The 
correlations between these excitations are anomalous 
and show up as interaction-dependent nonuniversal 
power-laws in many physical quantities where those 
of ordinary metals are characterized by universal 
(interaction independent) powers.

These properties are generic for one-dimensional Fermi 
liquids but particularly prominent in a one dimensional model 
of interacting fermions proposed by Tomonaga (Tomonaga 1950), 
Luttinger (Luttinger 1963) and solved exactly by Mattis and 
Lieb (Mattis and Lieb 1965). All correlation functions of 
the Luttinger model can be computed exactly, so that one 
has direct access to all physical properties of interest.
The notion of a Luttinger liquid was given by Haldane 
(Haldane 1981) to describe the universal low-energy 
properties of gapless one dimensional quantum systems, 
emphasizing that an asymptotic ($\omega \rightarrow 0, 
q \rightarrow 0$) description can be based on the Luttinger 
model in much the same way as the Fermi liquid theory
in three dimension is based on the free Fermi gas. The basic 
ideas and procedures had been discussed earlier by Efetov and 
Larkin (Efetov and Larkin 1975) but passed largely unnoticed.

\subsection{Thermodynamic Bethe Ansatz}

The first approach to extend Bethe Ansazt to finite 
temperatures was developed by Yang and Yang (Yang and Yang 
1969; Yang 1970) for the one dimensional boson gas with 
$\delta$-function interaction. The method is based on the 
analysis of the spin wave spectrum of the Bethe Ansatz equations. 
The first analysis of this spectrum took into account only real roots 
of the Bethe Ansatz equations (Cloiseaux and Pearson 1962;
Griffiths 1964). But in general the roots are complex, as 
first observed by Katsura (Katsura 1965; Ovchinnikov 1969),
which give bound states and may be grouped in various ``strings". 
This ``string" method was successfully implemented in all cases 
where the ground state contains bound pairs, such
as: {\it i}) the $\vert \Delta \vert = 1$ Heisenberg 
spin-chain (Takahashi 1971a).  
{\it ii}) the $\vert \Delta \vert \ge 1$ Heisenberg-Ising
(XXZ) spin-chain (Gaudin 1971; Takahashi and Suzuki 1972)
(extended also for general XYZ model (Takahashi and Suzuki 1972)). 
{\it iii}) the  1D attractive $\delta$-function interacting 
electron systems (Takahashi 1971b; Schlottmann 1993). 
Hence also for the $U < 0$ Hubbard model and half-filled 
$U > 0$ Hubbard model (Takahashi 1972); and {\it iv}) 
models with effective one dimensional attractive 
$\delta$-function interactions such as the Kondo model 
(Filyov, Tsvelick and Wiegmann 1981; Andrei and Lowenstein 1982;
Tsvelick and Wiegmann 1983a), the Coqblin-Schrieffer 
model (Tsvelick and Wiegmann 1982a;  Rasul 1982; 
Tsvelick and Wiegmann 1983a; Andrei, Furuya and Lowenstein 
1983) and the nondegenerate Anderson model with infinite $U$
(Filyov, Tsvelick and Wiegmann 1982; Tsvelick and Wiegmann 
1982b; Kawakami and Okiji 1982; Tsvelick and Wiegmann 1983a; 
1983b, 1983c; Okiji and N. Kawakami 1983; Schlottmann 1984). 

The string (or Thirring) hypothesis is valid 
for models exhibiting Luttinger liquid ground state properties,
repulsive $\delta$-function interacting
models (Takahashi 1971b, 1972; Lai, 1971, 1973), 
and the partially filled $U > 0$ Hubbard model
(Takahashi 1972). As such, the same properties apply. 
The Bethe Ansatz equations of the Hubbard model have three types 
of roots: {\it i}) real $k$, {\it ii}) $k$ and $\Lambda$ string
solutions and {\it iii}) pure $\Lambda$ strings. The existance
of the real $k$ roots shows that even at finite temperature 
pure holon type excitations, and hence spin - charge separation,
exist. For more details, see Korepin, Bogoliubov and Izergin
(1993). However, even in cases where the Bethe Ansatz method 
is applicable, correlations functions cannot be calculated and 
other methods, i.e. conformal field theory or bosonization has 
to be implemented. The later is our main focus. 

\subsection{Finite Temperature Formalism}

The static properties of an equilibrium system at given 
temperature can be ascertained by deriving the partition 
function of the system, giving standard thermodynamic properties, 
(Garrod 1995). To calculate the dynamical properties of systems 
at finite temperature we must adopt a formalism that includes the 
statistical mechanics of a system for a given temperature.

The finite temperature expectation values of a field theory 
are no longer ground state expectations, as they are for 
theories at $T = 0$, but expectations taken over an ensemble 
of states.  The operator corresponding to the observable must 
be averaged over an ensemble of basis states, with the weighting 
of each state determined by its energy.  Mathematically this 
is expressed by,
\begin{equation}
\langle \hat{A} \rangle_{\beta} = \frac{1}{\cal{Z}} {\rm{Tr}} \; 
e^{-\beta \hat{H}} \hat{A} \; , 
\end{equation}
where ${\rm{Tr}}$ is the trace over the basis states, 
and ${\cal{Z}}$ is the partition function ${\cal{Z}} = 
{\rm{Tr}} \; e^{-\beta \hat{H}}$.

A number of finite temperature formalisms have been developed, 
mainly relying on a mathematical correspondence between quantum 
field theories in D dimensions, and the statistical mechanics 
of systems in D+1 dimensions.

The simplest finite temperature formalism is the imaginary time, 
or Matsubara formalism.  In this formalism the equilibrium 
statistical mechanics of the system is equivalent to an 
imaginary time evolution.  The solutions can then be 
determined by standard methods using an ``imaginary time'' 
variable $\tau = it$, where $\tau$ is an inverse temperature.  
Unfortunately, this leads to the loss of any information about 
the dynamics of the system, as the time variable has been 
replaced by the temperature.  Therefore, only static 
equilibrium properties of the system can be analysed.  
Dynamic properties, such as spin-charge separation, are 
no longer observable in this formalism.

Dynamic finite temperature correlation functions may be 
calculated from the results of the Matsubara method, however, 
by using the spectral density function (Inkson 1984).  
The ``real time'' correlation functions are determined by 
analytically continuing the spectral density function.  
From the Matsubara method the spectral density function 
is known at infinitely many points on the imaginary line, 
the Matsubara frequencies $i\omega_n$.  If the complete 
spectral density function can be obtained, then the real 
time correlation functions can be given for arbitrary 
time and temperature.

Using the bosonization technique, the trace over a basis 
may be done using a basis of boson number states.  A 
direct evaluation of the expectation values may thus 
be obtained easily, without the use of Matsubara 
frequencies and analytic continuation.  Properties 
such as spin-charge separation may then easily be 
studied at finite temperature.

Hereafter, we review the bosonization of certain one dimensional 
models, and calculate the correlation functions, and some 
Green's functions, for these one dimensional models at finite 
temperature.

\section{Bosonization}

Bosonization of a quantum field theory describes a method by 
which fermionic operators in the theory, obeying anti-commutation 
relations, are replaced by bosonic operators obeying commutation 
relations.

Replacement of one field theory with another would appear to 
merely replace the problem of solving the original fermionic 
field theory with the problem of solving a bosonic field theory.  
The usefulness of this technique was realised when it was 
discovered that certain one dimensional {\sl interacting} 
fermionic field theories, were equivalent to {\sl non-interacting} 
bosonic field theories.  While the solution to interacting field 
theories is often difficult, and generally only perturbative 
methods are available to obtain a solution, the solution to a 
non-interacting field theory is well known and exact (Abrikosov, 
Gorkov and Dzyaloshinski 1963; Inkson 1984). 

Given the exact solution to the bosonic field theory, the 
properties of the fermionic theory can be calculated by use 
of a ``bosonization dictionary''.  For example, correlation 
functions in terms of fermionic operators can be re-written 
as expectations for bosonic operators, and solved.

Bosonization may be derived by defining a set of boson fields 
$\phi(x)$, and their conjugate momenta fields $\Pi(x)$, with 
given commutation relations, and then determining the commutation 
relations and Green's functions of the exponentials of these fields.  
The Green's functions and commutation relations are fermionic in 
nature, and an identity between the boson fields, $\phi(x)$ and 
$\Pi(x)$, and fermion fields $\psi(x)$ is made 
(Shankar 1995; Gul{\'a}csi 1997). 

Field theoretical bosonization evolved out of solutions to models 
in the area of high energy physics.  Indeed, many of the models 
used in the particle physics arena have a direct correspondence 
to models used in condensed matter physics.  Some relevant 
examples of this correspondence shall be pointed out later on.

\subsection{Bosonization Via Fock Space Identification}

Constructive bosonization, in terms of a Fock space identification 
of operators, gives a simpler, more transparent derivation of the 
boson representation of fermion operators.  An explicit picture of 
the physics corresponding to the boson representation may be 
obtained directly from the formalism. We adopt this procedure 
hereafter, not just because of its simplicity, but mainly because
it allows a straightforward extension to finite temperature. 

In this section we derive the Fock space identity between fermions 
and boson coherent states, closely following the pedagogical article 
of Sch{\"o}nhammer and Meden (1996); Delft and Sch\"{o}ller (1998).  
From this derivation we obtain the bosonization dictionary we 
require for the solution of the models included in later chapters.

\subsubsection{Requirements for Bosonization}

The fermionic system must be expressible by a countable set of 
operators, with unbounded spectrum, that obey fermionic 
anti-commutation relations.  That is, operators $\hat{c}_{k,\sigma}$ 
with $k\in [ -\infty , \infty ]$, $\sigma$ a species identifying 
label, and,
\begin{equation}
\{ \hat{c}_{k'\sigma'}, \hat{c}^{\dag}_{k\sigma} \} = 
\delta_{kk'}\delta_{\sigma \sigma'} \; . 
\end{equation}
The species identifying label can be just the electron spin 
$\sigma = \{ \uparrow, \downarrow \}$, or in the Luttinger model 
the spin combined with the left moving $L$ and right moving $R$ 
species labels.

The treatment of fermions on different sections of the Fermi 
surface as different species can be used to generalize this 
method to higher dimensions.  The success of this approach in 
one dimension hinges on the assumption that there is no mixing 
of fermion species, due to the large $2k_F$ jump in momentum 
between the points of the Fermi surface.

At this point there are no requirements on the type of Hamiltonian 
of the system, as the operator identity is a property of the Fock 
space.

\subsubsection{The Vacuum State and Normal Ordering}

In order to regularize all the expressions of operators within 
the theory, a zero particle vacuum state $|0\rangle$ for the 
fermions is introduced.  This vacuum state is then used to define 
normal orderings of all the operators.  The vacuum state is defined 
by,
\begin{eqnarray}
\hat{c}_{k\sigma}|0\rangle &= 0 \; ; 
\qquad \text{for } k>0 \; , 
\nonumber \\
\hat{c}^{\dag}_{k\sigma} |0\rangle &= 0 \; ; 
\qquad \text{for } k\leq 0 \; . 
\nonumber
\end{eqnarray}

Hence, for any combination of the fermion operators we may 
define the normal ordering of such operators by, 
$:\hat{A}\hat{B}\hat{C}: \; = \; \hat{A}\hat{B}\hat{C} -
\langle 0|\hat{A}\hat{B}\hat{C} |0\rangle$. 

\subsubsection{Defining the Boson Operators}

The boson operators $\hat{b}^{\dag}_{q\sigma}$ and 
$\hat{b}_{q\sigma}$ are defined on the Fock space as linear 
combinations of particle-hole excitations,
\begin{equation}
\hat{b}^{\dag}_{q\sigma} = \sqrt{\frac{2\pi}{Lq}} 
\sum_{k=-\infty}^{\infty} \hat{c}^{\dag}_{k+q,\sigma}
\hat{c}^{}_{k\sigma} \; , \; \; \; 
\hat{b}^{}_{q\sigma} = \sqrt{\frac{2\pi}{Lq}} 
\sum_{k=-\infty}^{\infty} \hat{c}^{\dag}_{k-q,\sigma}
\hat{c}^{}_{k\sigma} \; , 
\nonumber
\end{equation}
where the momentum $q=\frac{2\pi n}{L}$ for $n$ a positive integer.
The boson operators are proportional to the $q \neq 0$ Fourier 
components of the fermion density $\hat{\psi}^{\dag}_{\sigma}(x)
\hat{\psi}^{}_{\sigma}(x)$, with the $q=0$ component related to 
the fermion number operator $\hat{N}$.

The operators obey the bosonic commutation relations,
\begin{equation}
[ \hat{b}^{\dag}_{q,\sigma}, \hat{b}^{\dag}_{q',\sigma'} ] = 
[ \hat{b}^{}_{q,\sigma}, \hat{b}^{}_{q',\sigma'} ] = 0 \; , \; \; \; 
[ \hat{b}^{}_{q,\sigma}, \hat{b}^{\dag}_{q',\sigma'} ] = 
\delta_{qq'} \delta_{\sigma \sigma'} \; , 
\nonumber
\end{equation}
where the last equality requires normal ordering and the existence 
of the negative $k$ states,
\begin{eqnarray}
& \hat{b}^{}_{q\sigma} \hat{b}^{\dag}_{q'\sigma'} - 
\hat{b}^{\dag}_{q'\sigma'} \hat{b}^{}_{q\sigma} 
= \frac{2\pi}{Lq} \delta_{\sigma \sigma'} \sum_{k} 
\left[ \hat{c}_{k+q-q',\sigma}^{\dag} \hat{c}^{}_{k,\sigma} 
- \hat{c}_{k+q,\sigma}^{\dag} \hat{c}^{}_{k+q',\sigma} \right] 
\nonumber \\
& = \frac{2\pi}{Lq} \delta_{\sigma \sigma'} \delta_{qq'} 
\sum_{k} \left[ \langle 0 | \hat{c}_{k\sigma}^{\dag} 
\hat{c}^{}_{k\sigma}|0 \rangle - 
\langle 0 | \hat{c}_{k+q,\sigma}^{\dag} 
\hat{c}^{}_{k+q,\sigma} | 0 \rangle \right] 
= \delta_{\sigma \sigma'}  \delta_{qq'} \; . 
\nonumber
\end{eqnarray}

\subsubsection{Partitioning the Fock Space}

The fermionic Fock space can be interpreted as a direct sum 
of subspaces, each of definite particle number $N$.  Excitations 
within these subspaces are then bosonic particle-hole excitations, 
with the $\hat{b}^{}_q$ and $\hat{b}^{\dag}_q$ operators acting as 
annihilation and creation operators, respectively.  As first 
noted by Overhauser (1965), the bosonic particle-hole 
excitations form a complete basis of states for each $N$ particle 
subspace.

The operators that connect the subspaces together are the unitary 
Klein operators $\hat{F}^{\dag}$ and $\hat{F}^{}$, which increase and 
decrease the total particle number by one, respectively.  In this 
way, the Klein operators act as raising and lowering operators 
with respect to the fermion number operator.  They obey the 
consequent commutation relations,
\begin{equation}
[ \hat{N}^{}_{\sigma} , \hat{F}^{\dag}_{\sigma'} ] = 
\delta_{\sigma \sigma'} \hat{F}^{\dag}_{\sigma'} \; , \; \; \; 
[ \hat{N}^{}_{\sigma} , \hat{F}^{}_{\sigma'} ] = 
- \delta_{\sigma \sigma'} \hat{F}^{}_{\sigma'} \; . 
\nonumber \\
\end{equation}

The operators were first constructed in terms of the bare 
fermions by Haldane (1979), and an overview of their properties 
given by Haldane (1981); Delft and Sch{\"o}ller (1998).
For the purposes of this paper we simply describe remaining 
commutation relations obeyed by the Klein operators, which are,
\begin{eqnarray}
& \{ \hat{F}^{\dag}_{\sigma}, \hat{F}^{}_{\sigma'} \} = 
2 \delta_{\sigma \sigma'} \; , \; \; \; 
\{ \hat{F}^{\dag}_{\sigma}, \hat{F}^{\dag}_{\sigma'} \} = 
\{ \hat{F}^{}_{\sigma}, \hat{F}^{}_{\sigma'} \} = 0 \; , 
\nonumber \\
& [ \hat{b}^{\dag}_{q\sigma} , \hat{F}^{\dag}_{\sigma'} ] = 
[ \hat{b}^{}_{q\sigma} , \hat{F}^{\dag}_{\sigma'} ] = 
[ \hat{b}^{\dag}_{q\sigma} , \hat{F}^{}_{\sigma'} ] = 
[ \hat{b}^{}_{q\sigma} , \hat{F}^{}_{\sigma'} ] = 0 \; . 
\label{klein_commute}
\end{eqnarray}
We also note that because the Klein operators are unitary, 
we have, $\hat{F}^{\dag}_{\sigma} \hat{F}^{}_{\sigma} = 
\hat{F}^{}_{\sigma} \hat{F}^{\dag}_{\sigma} = 1$.

\subsubsection{Fermion Operators as Coherent States}

The fermion field operators $\psi^{}_{\sigma}(x)$ and 
$\psi^{\dag}_{\sigma}(x)$ are defined in terms of the Fourier 
components of the $\hat{c}^{}_{k\sigma}$ and 
$\hat{c}^{\dag}_{k\sigma}$ operators, respectively,
\begin{equation}
\psi_{\sigma}(x) = \sqrt{\frac{2\pi}{L}} \sum_{k=-\infty}^{\infty} 
e^{-ikx} \hat{c}^{}_{k\sigma} \; , \; \; \; 
\psi^{\dag}_{\sigma}(x) = \sqrt{\frac{2\pi}{L}}
\sum_{k=-\infty}^{\infty} e^{ikx} \hat{c}^{\dag}_{k\sigma} \; . 
\end{equation}

The fundamental relation that makes bosonization work is that 
the state produced by the fermion operator 
$\hat{\psi}^{}_{\sigma}(x)$ on an $N$ particle ground state 
$|N_0\rangle$ is an eigenstate of the boson operator 
$\hat{b}^{}_{q,\sigma}$.  This may be derived from the 
commutation relations,
\begin{equation}
[ \hat{b}^{}_{q\sigma} , \hat{\psi}^{}_{\sigma'}(x) ] = 
-\delta_{\sigma \sigma'} \sqrt{\frac{2\pi}{Lq}}e^{iqx} 
\hat{\psi}^{}_{\sigma'}(x) \; , \; \; \; 
[ \hat{b}^{\dag}_{q\sigma} , \hat{\psi}^{}_{\sigma'}(x) ] = 
-\delta_{\sigma \sigma'} \sqrt{\frac{2\pi}{Lq}}e^{-iqx} 
\hat{\psi}^{}_{\sigma'}(x) \; , 
\end{equation}
and the fact that any $N$ particle ground state $|N_0\rangle$ 
vanishes when operated on by $\hat{b}_{q\sigma}$.  This allows 
the representation of the fermion operator 
$\hat{\psi}^{}_{\sigma}(x)$ acting on any $N$ particle ground 
state as a coherent state representation of boson operators.  
Thus,
\begin{equation}
\hat{\psi}^{}_{\sigma}(x) | N_0 \rangle = \hat{F}^{}_{\sigma} 
\lambda_{\sigma}(x) \exp \left( -\sum_{q>0} 
\sqrt{\frac{2\pi}{Lq}}e^{iqx} \hat{b}^{\dag}_{q\sigma} \right) 
| N_0 \rangle \; , 
\end{equation}
where the Klein operator $\hat{F}^{}_{\sigma}$ is required to 
reduce the particle number, and $\lambda_{\sigma}(x)$ is a phase 
operator,
\begin{equation}
\lambda_{\sigma}(x) = \sqrt{\frac{2\pi}{L}} e^{-i \frac{2 \pi }{L} 
( \hat{N}-\frac{1}{2}\delta_B ) x } \; . 
\end{equation}

The boson field operators $\varphi_{\sigma}(x)$ and 
$\varphi^{\dag}_{\sigma}(x)$, are introduced to simplify the 
notation, and are simply,
\begin{equation}
\varphi_{\sigma}(x) = -\sum_{q>0} \sqrt{\frac{2\pi}{Lq}} 
e^{iqx} \hat{b}_{q\sigma} \; , \; \; \; 
\varphi_{\sigma}^{\dag}(x) = -\sum_{q>0} \sqrt{\frac{2\pi}{Lq}} 
e^{-iqx} \hat{b}^{\dag}_{q\sigma} \; . 
\end{equation}
and allow us to write $\hat{\psi}_{\sigma}(x)|N_0 \rangle = 
\hat{F}_{\sigma} \lambda_{\sigma}(x) e^{\varphi^{\dag}_{\sigma}(x)} 
| N_0 \rangle$.

\subsubsection{The Boson Representation of $\hat{\psi}_{\sigma}(x)$}

Bosonization is only useful because properties of the fermionic 
system can be described in terms of certain boson fields, whose 
properties are simpler to calculate than those of the original 
fermion fields.  The correlation functions, and hence Green's 
functions, of the fermionic system contain combinations of 
single particle fermion operators, hence we need to express 
these operators in terms of a set of boson fields.

In the previous section we showed that the fermion field 
$\hat{\psi}^{}_{\sigma}(x)$ can be expressed as a coherent 
state of boson operators.  The generalization of this boson 
representation to an arbitrary state of the Fock space is 
surprisingly straight forward.

As the boson excitations form a complete basis of any $N$ 
particle subspace we can write an arbitrary state $|\Psi\rangle$ 
as some set of boson excitations above an $N$ particle ground 
state, that is, $|\Psi\rangle = f( \{ \hat{b}^{\dag}_{q\sigma} \} ) 
| N_0\rangle $.

Using the commutation relations for $\hat{\psi}_{\sigma}(x)$ 
and $\hat{b}^{\dag}_{q\sigma}$, and the relation,
\begin{equation}
e^{-\varphi_{\sigma}(x)} f( \{ \hat{b}^{\dag}_{q'\sigma'} \} ) 
e^{\varphi_{\sigma}(x)} = f( \{ \hat{b}^{\dag}_{q'\sigma'} + 
\delta_{\sigma \sigma'} \sqrt{\frac{2\pi}{Lq}}e^{iqx} \} )
\end{equation}
we can derive the bosonization identity,
\begin{equation}
\hat{\psi}_{\sigma}(x)|\Psi \rangle = \hat{F}_{\sigma} 
\lambda_{\sigma}(x) e^{\varphi^{\dag}_{\sigma}(x)} 
e^{-\varphi_{\sigma}(x)} |\Psi \rangle \; . 
\label{boson_ident}
\end{equation}
The boson representation of the fermion fields allows us to 
obtain fermion expectation values in terms of the expectation 
values of the exponentials of the boson fields.
In similar treatments, (Emery 1979; Mahan 1981;  
Delft and Sch{\"o}ller 1998), the boson operators 
$\hat{b}^{}_{q\sigma}$ and $\hat{b}^{\dag}_{q\sigma}$, are 
given factors of $-i$ and $i$, respectively, which changes 
the exponentials in Eq.\ (\ref{boson_ident}) from $\pm$ to $-i$ 
for both terms.

\subsection{Bosonization of Hamiltonians}

The definition of the boson operators as bilinear combinations 
of fermion operators would suggest that the two particle 
interaction terms of a fermionic Hamiltonian, which may be 
expressed as an effective density-density interaction, may be 
rewritten in terms of a sum of quadratic combinations of boson 
operators.

If the kinetic term of the fermion Hamiltonian, which in the 
case of a Fermi gas, only contains a single bilinear combination 
of fermion operators, was also expressible in terms of quadratic 
combinations of boson operators, then the model may easily be 
diagonalised in terms of the boson operators.

By commuting the boson annihilation operator $\hat{b}^{}_{q\sigma}$ 
with the kinetic term of a fermion Hamiltonian of the form, 
$\hat{H}_0 = \sum_{k,\sigma} \epsilon (k) \; 
\hat{c}_{k\sigma}^{\dag} \hat{c}^{}_{k\sigma}$,
\begin{equation}
[ \hat{H}_0 , \hat{b}^{}_{q\sigma} ] = 
\sum_k (\epsilon (k-q)-\epsilon (k) ) \;  
\hat{c}_{k-q,\sigma}^{\dag} \hat{c}^{}_{k\sigma} \; , 
\label{linear_require}
\end{equation}
we find, if the quantity $\epsilon (k-q)-\epsilon (k)$ is 
independent of $k$, the commutation relation is that of a 
lowering operator $\hat{b}^{}_{q\sigma}$.  The the kinetic 
term of the Hamiltonian may then be expressed as a bilinear 
combination of the operator $\hat{b}_{q\sigma}$ and its adjoint.  
This identification of the boson and fermion Hamiltonians is 
known as Kr\"{o}nig's identity (Kr{\"o}nig 1935).

By examination we can see, for the term $\epsilon (k-q) - 
\epsilon (k)$ to be independent of $k$, the spectrum must be 
linear in $k$, that is, $\epsilon(k) = A k +B$, for some 
constants $A$ and $B$.  For this reason, in the next chapter, 
we begin our examination of fermions in one dimension with a 
gas of fermions with an infinite linear dispersion.

\section{Fermions with Linear Dispersion}

These fermions will have an infinite linear dispersion 
$\epsilon(k)=v_F(k - k_F)$,  
for $v_F$ the Fermi velocity, and $k_F$ the Fermi momentum. 
The model has an infinite Fermi sea of positron states.  
However, these states do not contribute to the low energy 
physics of the model, and therefore the physics is believed 
to be descriptive of the low energy sector of more realistic 
one dimensional models. 
Restriction of the particles to movement in one direction 
also has physical realisations.  The transport in quantum 
Hall edge states is an example of one such occurrence.

The construction of a model with infinite linear dispersion 
provides an exact bosonization of the model with a non-interacting 
boson Hamiltonian. All the properties of the fermionic system may 
then be obtained by direct calculation, using the boson 
representation. As this model is critical at zero temperature, 
the scale and Lorentz invariance of the system allows it to be 
solved using conformal field theory (Christe and Henkel 1993).  
This is possible due to the Fermi velocity acting as a Lorentz 
invariant velocity equivalent to the speed of light in conformal 
field theory models.  Models which contain excitations with more 
than a single velocity may still be solved using a product of 
Virasoro algebras and a dressed charge matrix (Frahm and 
Korepin 1990).

\subsection{The Non-Interacting Spinless Model}

In the non-interacting case, 
$\hat{H}_0 = v_F \: \sum_{k=-\infty}^{\infty} \: (k - k_{F}) \:  
\hat{c}^{\dag}_k \hat{c}^{}_k$, the direct solution of the fermion 
model requires less work than the bosonization method.  However, 
the bosonization gives some insight into the structure of one 
dimensional fermion systems.  The bosonization solution also 
extends easily to the interacting case, unlike the direct 
fermionic solution. Being on a lattice we also have: 
\begin{equation}
k = \frac{2\pi}{L}\left( n_k -\frac{1}{2} \delta_B \right) \; , 
\end{equation}
$n_k \in {\rm{Z}}$, $\delta_B \in [0,2)$ a parameter that determines 
the boundary conditions, and $L$ the length of the system. 

Defining the bosons, for $q>0$,
\begin{equation}
\hat{b}^{\dag}_q = \sqrt{\frac{2\pi}{Lq}} \sum_{k} 
\hat{c}^{\dag}_{k+q} \hat{c}_k \; , \; \; \; 
\hat{b}^{}_q = \sqrt{\frac{2\pi}{Lq}} \sum_{k} 
\hat{c}^{\dag}_{k-q} \hat{c}_k \; , 
\end{equation}
we then commute the boson operators with the Hamiltonian, to find,
\begin{eqnarray}
[ \hat{H}_0 ,\hat{b}^{}_q ] &= \sqrt{\frac{2\pi}{Lq}} v_F 
\sum_{k'} \left[ (k'-q-k_F) \hat{c}^{\dag}_{k'-q} 
\hat{c}_{k'} - (k'-k_F) \hat{c}^{\dag}_{k'-q} \hat{c}^{}_{k'} \right] 
\nonumber \\
&= \sqrt{\frac{2\pi}{Lq}} v_F \sum_{k'} (-q) 
\hat{c}^{\dag}_{k'-q} \hat{c}^{}_{k'} = -v_Fq \; \hat{b}^{}_q \; . 
\end{eqnarray}
Using Kr\"{o}nig's identity (Kr{\"o}nig 1935) we can rewrite 
the Hamiltonian as a sum over bi-linear boson operators,
\begin{equation}
\hat{H}_0 = v_F \sum_q q \; \hat{b}^{\dag}_q \hat{b}^{}_q + 
\hat{C} \; , 
\end{equation}
where $\hat{C}$ is an operator that depends on the energies 
of the bosonic ground states. To determine the operator 
$\hat{C}$ we apply the fermion Hamiltonian to an arbitrary 
ground state of $N$ particles,
\begin{equation}
\hat{H}_0 | N_0 \rangle = \frac{2\pi}{L}v_F \sum_n n-\frac{1}{2}
\delta_B | N_0 \rangle = \frac{\pi v_F}{L} N(N+1-\delta_B) 
|N_0 \rangle \; . 
\end{equation}
To obtain this result we must include in the bosonized 
Hamiltonian the same combination of fermion number operators, 
which must also be normal ordered, as $N$ does not include 
states below the zero particle ground state energy.  With the 
zero particle ground state denoted by $| 0 \rangle$, and defined 
by $\hat{c}^{}_k |0\rangle = 0$ for $k>0$, and 
$\hat{c}^{\dag}_k |0\rangle = 0$ for $k\leq 0$, we define the 
fermion number operators by,
\begin{equation}
\hat{N} = \sum_k \left[ \hat{c}^{\dag}_k \hat{c}^{}_k - 
\langle 0 | \hat{c}^{\dag}_k \hat{c}^{}_k | 0\rangle \right] \; . 
\end{equation}

Hence, we may write the full boson Hamiltonian as,
\begin{equation}
\hat{H}_0 = v_F \sum_q q\; \hat{b}^{\dag}_q \hat{b}^{}_q + 
\frac{\pi v_F}{L} \hat{N}(\hat{N}+1-\delta_B) \; . 
\end{equation}
This Hamiltonian describes a system of free bosons with 
energies given by $\epsilon_q = v_F q$ for $q > 0$, with 
an additional term that gives the bosonic ground state 
energies for each fixed particle number subspace of the 
fermion Hilbert space.

\subsubsection{Distribution Function for the Bosons}
\label{distrib_func}

In the finite temperature calculation of expectation values, 
we must average over an ensemble of states.  We must therefore 
check that the extra term in the Hamiltonian, which gives the 
energies of the ``ground states'' for the bosons, does not 
change the distribution function of these bosons at finite 
temperature.

To begin with we simply calculate the partition function, 
${\cal{Z}}$, where,
\begin{eqnarray}
&{\cal{Z}} = \sum_N \sum_{ \{ n_q \} } \langle N,\{ n_q \} | 
e^{-\beta \hat{H}_0 } | N,\{ n_q \} \rangle 
\nonumber \\
&= \sum_N e^{-\beta \frac{\pi v_F}{L}N(N+1-\delta_B)} 
\sum_{ \{ n_q \} } \langle N,\{ n_q \} | 
e^{-\beta \sum_q \epsilon_q n_q } | N,\{ n_q \} \rangle 
\nonumber \\
&= \sum_N e^{-\beta \frac{\pi v_F}{L}N(N+1-\delta_B)} 
\prod_q \sum_{n=0}^{\infty} e^{-\beta \epsilon_q  n } 
\nonumber \\
&= \sum_N e^{-\beta \frac{\pi v_F}{L}N(N+1-\delta_B)} \: / \: 
\prod_q ( 1 - e^{-\beta \epsilon_q} ) \; , 
\end{eqnarray}
the third line is obtained from the second, by first expressing 
the sum in the exponential as a product of exponentials, and 
secondly summing over the occupation numbers.  The series is 
summed by using $\sum_{n=0}^{\infty} x^n = (1-x)^{-1}$. Further, 
we note that the chemical potential for the bosons vanishes.

Calculating the boson distribution function, we obtain,
\begin{eqnarray}
& \langle \hat{b}^{\dag}_q\hat{b}_q \rangle_{\beta} 
= \frac{1}{\cal{Z}} \sum_N \sum_{ \{ n_{q'} \} } 
\langle N,\{ n_{q'} \} | e^{-\beta  \hat{H}_0} \, 
\hat{b}^{\dag}_q \hat{b}^{}_q | N,\{ n_{q'} \} \rangle 
\nonumber \\
&= \frac{1}{\cal{Z}} 
\sum_N e^{-\beta \frac{\pi v_F}{L}N(N+1-\delta_B)} 
\sum_{ \{ n_{q'} \} } \langle N,\{ n_{q'} \} | e^{-\beta \sum_{q'} 
\epsilon_{q'}  n_{q'}} \hat{b}^{\dag}_q \hat{b}^{}_q | N,\{ n_{q'} \} 
\rangle \nonumber \\
&= \frac{1}{\cal{Z}} 
\sum_N e^{-\beta \frac{\pi v_F}{L}N(N+1-\delta_B)} 
\sum_{ \{ n_{q'} \} } \prod_{q'} e^{-\beta \epsilon_{q'} n_{q'} } 
\langle N,\{ n_{q'} \} | \hat{b}^{\dag}_q \hat{b}^{}_q | 
N, \{ n_{q'} \} \rangle \nonumber \\
&= \sum_{n=0}^{\infty} n e^{-\beta \epsilon_{q} n } 
\left( 1 - e^{-\beta \epsilon_{q}} \right) \sum_{ \{ n_{q'} \} } 
\prod_{q'\neq q} e^{-\beta \epsilon_{q'} n_{q'} } 
\left( 1 - e^{-\beta \epsilon_{q'}} \right) 
= 1 / ( e^{\beta \epsilon_q}-1 ) \; . 
\end{eqnarray}
Above we use the series summation $\sum_{n=0}^{\infty} n 
x^n = x(1-x)^{-2}$, and note that the sum over states of 
the product, in the second last line, is equal to one.

The terms introduced by the fermion number operator in the 
Hamiltonian cancel, leaving a free boson distribution function.  
The lack of dependence of the distribution function on the boson 
ground states can be understood physically, in that, each 
particle-hole excitation only depends on the difference in 
energy between the hole and particle, and not on the energies 
of the particle and hole individually.

The partition function, however, relates to the underlying 
fermion system, and hence is not equal to the partition 
function of a gas of bosons, and will give fermionic 
thermodynamic properties.  The equality of the fermion 
and boson partition functions is shown explicitly in 
the article by Haldane (1981), and is essential 
in proving the completeness of the boson states.

\subsubsection{Time Dependence of the Boson Fields}

For the non-interacting case, the time dependence of the 
boson operators is trivially obtained due to the 
non-interacting nature of the boson Hamiltonian.  
In deriving the time dependence of the boson fields 
we shall generalize the result for the case of a free 
boson Hamiltonian with arbitrary dispersion.  This will 
allow the use of these results in later sections.

The boson operators $\hat{b}^{}_q$ have a time dependence given by,
\begin{equation}
\hat{b}^{}_{q}(t) = e^{i\hat{H}t} \hat{b}^{}_{q} e^{-i\hat{H}t} = 
e^{-i\epsilon_q t} \hat{b}^{}_{q} \; , 
\end{equation}
where $\epsilon_q$ is the dispersion, which in the non-interacting 
case is simply $\epsilon_q = v_F q$.

We define the boson fields,
\begin{equation}
\varphi(x) = -\sum_q \sqrt{\frac{2\pi}{Lq}} e^{iqx} \hat{b}^{}_q
\; , \; \; \; 
\varphi^{\dag}(x) = -\sum_q \sqrt{\frac{2\pi}{Lq}} e^{-iqx} 
\hat{b}^{\dag}_q \; , 
\end{equation}
which when commuted with the free boson Hamiltonian, introduce a 
factor given by the dispersion $\epsilon_q$ for each 
$\hat{b}^{}_{q\sigma}$ operator in the sum,
\begin{equation}
[ \hat{H}_0 , \varphi(x) ] 
= -\sum_{q,q'} \epsilon_q \sqrt{\frac{2\pi}{Lq'}} e^{iq'x} 
[ \hat{b}^{\dag}_q \hat{b}_q , \hat{b}^{}_{q'} ] 
= \sum_{q} \epsilon_q \sqrt{\frac{2\pi}{Lq}} e^{iqx} 
\hat{b}^{}_{q} \; . 
\end{equation}
Thus, the time dependence of the field operators is easily 
obtained by moving the time dependent operators inside the sum, 
to obtain,
\begin{equation}
\varphi(x,t) = e^{i\hat{H}t} \hat{\varphi}(x) e^{-i\hat{H}t} 
= -\sum_{q} \sqrt{\frac{2\pi}{Lq}} e^{iqx} e^{-i\epsilon_q t} 
\hat{b}^{}_{q} \; , 
\end{equation}
which in the non-interacting case reduces to 
$\varphi(x,t) = \varphi(x-v_Ft)$.  Similarly, we find,
\begin{equation}
\varphi^{\dag}(x,t) = \sum_{q} \sqrt{\frac{2\pi}{Lq}} 
e^{-iqx} e^{i\epsilon_q t} \hat{b}^{\dag}_{q} \; 
\end{equation}
and, in the non-interacting case $\varphi^{\dag}(x,t) = 
\varphi^{\dag}(x-v_Ft)$.

Returning explicitly to the case $\epsilon_q = v_F q$, the 
simple time dependence of the $\varphi (x,t)$ boson fields 
may be shown to extend to the time dependence of the coherent 
state $e^{-\varphi(x,t)}=e^{-\varphi(x-v_Ft)}$.  Thus, the 
evolution in time of the exponentials of the boson fields 
results in a translation of the fields by $v_Ft$.

Using a different derivation we can show explicitly how the 
Hamiltonian acts on the boson field state as a generator of 
translations.  To begin, the commutators of the fields with 
the Hamiltonian may be rewritten in the form,
\begin{equation}
[ \hat{H}_0 , \varphi(x) ] = i v_F \frac{\partial \varphi(x)}
{\partial x} \; , \; \; \; 
[ \hat{H}_0 ,\varphi^{\dag}(x) ] = i v_F 
\frac{\partial\varphi^{\dag}(x)}{\partial x}  \; . 
\end{equation}
From this we can calculate the commutator of the Hamiltonian with 
the exponential of the $\varphi(x)$ fields, from which we obtain,
\begin{eqnarray}
& [ \hat{H}_0 , e^{-\varphi(x)} ] 
= e^{-\varphi(x)} \left( -[\hat{H}_0,\varphi(x)] + \frac{1}{2!}
[[\hat{H}_0 , \varphi(x) ], \varphi(x)] + ... \right) 
\nonumber \\
&= -e^{-\varphi(x)} [\hat{H}_0,\varphi(x)] 
= -i v_F e^{-\varphi(x)}\frac{\partial \varphi(x)}{\partial x} 
= i v_F \frac{\partial}{\partial x} e^{-\varphi(x)} \; , 
\end{eqnarray}
where the second line follows from the first because the boson 
field commutes with its derivatives. Applying the Baker-Hausdorff 
identity to the time evolved exponential of the boson field,
\begin{equation}
e^{i\hat{H}_0t}e^{-\varphi(x)}e^{-i\hat{H}_0t} 
= e^{-\varphi(x)} - v_F t \frac{\partial e^{-\varphi(x)}}
{\partial x} + \left( v_F t \right)^2 \frac{1}{2!}
\frac{\partial^2 e^{-\varphi(x)}}{\partial x^2} + ... 
= e^{-\varphi(x-v_Ft)} \; . 
\end{equation}
Similarly, we can show,
\begin{equation}
[ \hat{H}_0 , e^{\varphi^{\dag}(x)} ] = i v_F 
\frac{\partial}{\partial x} e^{\varphi^{\dag}(x)} 
\; , \; \; \; 
e^{i\hat{H}_0t}e^{\varphi^{\dag}(x)}e^{-i\hat{H}_0t} = 
e^{\varphi^{\dag}(x-v_Ft)} \; . 
\end{equation}

The dependence of the fields on the combination $x\pm v_Ft$, 
may also be determined from a Lorentz invariance of the Hamiltonian.
The velocity $v_F$ replaces the velocity of light as the fundamental 
velocity in the theory (S{\'e}n{\'e}chal 1999).

The Lorentz invariance remains even in the presence of delta 
function interactions, due to the separation and independence of 
the charge and spin sectors of the interacting model, and linear 
spectrum of the excitations.  In this case, there exist two 
independent Lorentz invariant systems with velocities $v_{\varrho}$ 
and $v_{\chi}$.

\subsubsection{The Klein Factors}

Here we calculate the commutators of the Klein factors 
$\hat{F}, \hat{F}^{\dag}$, with the phase factor $\lambda(x)$, 
and also the time dependence of the Klein factors with the free 
boson Hamiltonian.

First, we calculate the commutation relations between the Klein 
factors and the phase factor,
\begin{eqnarray}
& [\hat{F},\lambda(x)] = [ \hat{F}, \sqrt{\frac{2\pi}{L}}
e^{-i\frac{2\pi}{L}\left( \hat{N}-\frac{1}{2}\delta_B \right) x}] 
= \sqrt{\frac{2\pi}{L}}e^{i\frac{\pi}{L}\delta_B x} 
[ \hat{F}, e^{-i\frac{2\pi}{L}\hat{N}x}] 
\nonumber \\
&= \sqrt{\frac{2\pi}{L}}e^{i\frac{\pi}{L}\delta_B x} 
e^{-i\frac{2\pi}{L}\hat{N}x} \left( -i\frac{2\pi}{L}x
[\hat{F},\hat{N}] - \frac{1}{2!} \left( \frac{2\pi}{L}x \right)^2 
[[\hat{F},\hat{N}],\hat{N}] + ... \right) 
\nonumber \\
&= \sqrt{\frac{2\pi}{L}}e^{-i\frac{2\pi}{L} 
\left( \hat{N} + 1 - \frac{\delta_B}{2} \right)x} \hat{F} \; , 
\end{eqnarray}
which gives the relation,
\begin{equation}
\hat{F}\lambda(x) = \left( 1+e^{-i\frac{2\pi}{L}x}\right) 
\lambda(x) \hat{F} \; . 
\end{equation}

The time dependence of the Klein factors depends only on the 
number operator part of the free boson Hamiltonian.  Thus,
\begin{eqnarray}
&\hat{F}(t) = e^{i\frac{\pi v_F}{L} \hat{N}(\hat{N}+1-\delta_B)t} 
\hat{F} e^{-i\frac{\pi v_F}{L} \hat{N}(\hat{N}+1-\delta_B)t} 
\nonumber \\
&= e^{-i\frac{2\pi v_F}{L} \left( \hat{N}+1-\frac{\delta_B}{2} 
\right) t} \hat{F} 
= \sqrt{\frac{L}{2\pi}} e^{-i\frac{2\pi}{L}v_F t} \lambda(v_F t) 
\hat{F} \; . 
\end{eqnarray}

These phase factors have all previously been ignored in calculating 
expectation values of fermion operators (Mattis and Lieb 1965; 
Voit 1995; Delft and Sch{\"o}ller 1998), when taking the limit 
$L\rightarrow \infty$.  In doing so we are making the presumption 
that the system has a fixed number of particles $N$, and in the 
thermodynamic limit $N/L \rightarrow \rho$, for some constant 
$\rho$.

If we take the phase factor for a fixed $N$, we can see that the 
phase term $2\pi N / L$ is just the Fermi wave number $k_F$.  
Extrapolating this to the general contribution of the $q=0$ phase 
terms, we can now presume these to generate the chemical potential 
$\mu$ in the boson representation.  As has been shown in Section 
\ref{distrib_func}, the explicit chemical potential vanishes in the 
boson representation.  Our above reasoning shows that it is retained 
in the formalism, only it is now related to the $q=0$, or $\hat{N}$ 
operator, dependence.

To examine the finite size effects on this system, or if the 
particle number varies over the whole fermion Fock space, these 
terms must explicitly be included.  However, in reducing a one 
dimensional system to the linearised model we are only considering 
excitations close to the Fermi surface, which corresponds to 
obtaining the large distance behaviour of the correlation 
functions and Green's functions.  In this case we simply 
ignore the contribution of the Klein operators and phase 
factors, and add a term $e^{ik_F x}$ to account for the 
chemical potential in the correlation functions.

\subsubsection{Finite Temperature Green's Functions}

The correlation functions $\langle \psi(x,t) \psi^{\dag}(0,0) 
\rangle_{\beta}$ and $\langle \psi^{\dag}(0,0) \psi(x,t) 
\rangle_{\beta}$ are easily calculated directly from the time 
dependent momentum representation of the fermion operators, 
followed by Fourier transforming over $k$.  From this we obtain,
\begin{equation}
\langle \psi(x,t) \psi^{\dag}(0,0) \rangle_{\beta} = 
\int \frac{e^{ikx} e^{-i(\epsilon_k-\mu)t}}{1+e^{-\beta(\epsilon_k 
- \mu)}} \; dk = \frac{i\pi e^{ik_Fx}}{\beta v_F \sinh 
\frac{\pi (x-v_Ft)}{\beta v_F}} \; . 
\end{equation}

The retarded Green's function, $G_R^{\beta}(x,t) = -i\theta(t)
\langle [ \psi(x,t), \psi^{\dag}(0,0) ]_+ \rangle_{\beta} $, 
can be calculated easily for the free fermion Hamiltonian, and 
for inverse temperature $\beta$.  From this calculation we obtain,
\begin{eqnarray}
&G^{\beta}_R(x,t) = -i\theta(t)\frac{2\pi}{L} \sum_k 
\left[ \frac{e^{ikx}e^{-i(\epsilon_k -\mu)t}}{1+e^{-\beta(\epsilon_k 
- \mu)}} + \frac{e^{ikx}e^{-i(\epsilon_k-\mu )t}}
{1+e^{\beta(\epsilon_k - \mu)}} \right] 
\nonumber \\
&= -i\theta(t)\frac{2\pi}{L} \sum_k e^{ikx}e^{-i(\epsilon_k -\mu)t} 
= -i\theta(t)e^{ik_Fx}\delta(x-v_F t) \; . 
\end{eqnarray}

The calculation using bosonization is more complicated for this 
simple case, but the generalization to the interacting case is 
straight forward, whereas the fermionic case is not. By expressing 
the fermion operators $\hat{\psi}(x,t)$ in terms of the boson 
operators, we note that the Klein factors pick up a time dependence 
due to the number operators in the Hamiltonian, and the exponential 
of the boson operators obtains a time dependence from the bilinear 
boson term.  That is,
\begin{equation}
\hat{\psi}(x,t) = e^{i\hat{H}t}\hat{\psi}(x)e^{-i\hat{H}t} 
= \sqrt{\frac{L}{2\pi}} e^{-i\frac{2\pi}{L}v_F t} \lambda(v_Ft) 
\hat{F} \lambda(x) e^{\varphi^{\dag}(x-v_Ft)} e^{-\varphi(x-v_Ft)} \; .
\end{equation}

The correlation functions are now given by,
\begin{eqnarray}
&\langle \psi(x,t) \psi^{\dag}(0,0) \rangle_{\beta} 
= \left\langle e^{-i\frac{2\pi}{L}v_F t} \sqrt{\frac{L}{2\pi}} 
\lambda(v_Ft) \hat{F} \lambda(x) \lambda^{\dag}(0) \hat{F}^{\dag} 
e^{\varphi^{\dag}(x-v_Ft)} e^{-\varphi(x-v_Ft)} 
e^{-\varphi^{\dag}(0)} e^{\varphi(0)} \right\rangle_{\beta} 
\nonumber \\
&= \left( 1+e^{-i\frac{2\pi}{L}x} \right) e^{-i\frac{2\pi}{L}v_Ft} 
\; \left\langle \lambda(x+v_Ft)  
e^{\varphi^{\dag}(x-v_Ft)} e^{-\varphi(x-v_Ft)} 
e^{-\varphi^{\dag}(0)} e^{\varphi(0)} \right\rangle_{\beta} \; . 
\label{complete_correl}
\end{eqnarray}
The term involving the fermion number operator will introduce 
phase factors in the sum over the boson ground states, which 
will not necessarily cancel with the terms in the partition 
function, giving phase factors related to the chemical potential.  
We ignore these for now, and concentrate on the expectation of the 
boson operators.

As the time dependence may be included by the exchange of $x$ 
for $x-v_Ft$, from this point we only explicitly write the $x$ 
dependence.

The expectation of the four exponential boson fields can be 
simplified by commuting the exponentials to move all the boson 
creation operators to the left of the annihilation operators.  
The commutation introduces a term, which we denote by,
\begin{equation}
D(x,0) = e^{[ \varphi(x), \varphi^{\dag}(0) ] }
= \exp \left( \frac{2\pi}{L} \sum_q \frac{e^{iqx}}{q} \right) \; , 
\end{equation}
that diverges in the limit of large system size.  Therefore, this 
term must be included back into the expectation at a future stage 
of the calculation in order to obtain correct results.

The expectation of the exponentials of the boson fields can be 
solved simply by using a relation for expectation values of 
exponentials, where,
\begin{equation}
\left\langle \exp \left( \sum_q A_q \hat{b}^{\dag}_q \right) 
\exp \left( \sum_q B_q \hat{b}_q\right) \right\rangle = 
\exp \left( \sum_q A_q B_q (e^{\beta \omega_q}-1)^{-1} \right) \; , 
\label{expectation}
\end{equation}
in the case where the Hamiltonian is identical to that of a 
simple harmonic oscillator, so we can follow it's solution 
(Mermin 1966). Ignoring the preceding phase factors generated 
by the Klein and $\lambda$ terms, we may express the expectation 
of the boson fields in the form,
\begin{eqnarray}
& \langle \psi(x) \psi^{\dag}(0) \rangle_{\beta} \simeq D(x,0) 
\bigg\langle \exp \left( \sum_q \sqrt{\frac{2\pi}{Lq}} (1-e^{-iqx}) 
\hat{b}^{\dag}_q \right) 
\nonumber \\
& \times \exp \left( \sum_q \sqrt{\frac{2\pi}{Lq}} (e^{iqx}-1) 
\hat{b}_q \right) \bigg\rangle \; . 
\label{incomp_correl}
\end{eqnarray}
which gives the $A_q$ and $B_q$ terms in Eq.\ (\ref{expectation}).

In the continuum limit the sum in the exponential in 
Eq.\ (\ref{expectation}) becomes an integral, and the correlation 
functions can be evaluated explicitly by solving the integrals,
\begin{equation}
2 \int_0^{\infty} \frac{\cos qx - 1}{q(e^{\beta v_F q} - 1)} dq + 
\int_0^{\infty} \frac{dq}{q} e^{iqx} \; , 
\end{equation}
where the first integral gives the contribution of the expectation 
of the fields, and the second integral comes from the $D(x,0)$ term 
generated by the commutator.

By combining the two terms we may be solve the integral explicitly, 
using the integral representation of the logarithm of the gamma 
function given by Ryzhik and Gradstein (1965), and we obtain the 
expression,
\begin{equation}
\int_0^{\infty} \frac{2}{q} \left[ \frac{\cos qx - 1}{e^{\beta v_F q}
- 1} + \frac{e^{iqx}}{2} \right] dq = 
\ln \left[ \Gamma\left( -\frac{ix}{\beta v_F} \right) 
\Gamma\left( 1+\frac{ix}{\beta v_F} \right) \right] + Q \; , 
\label{integral}
\end{equation}
where the divergent constant $Q$, is given by,
\begin{equation}
Q = \int_0^{\infty} \frac{e^{-\beta v_F q}}{q} \, dq \; . 
\end{equation}

The divergent constant $Q$ does not depend on $x$ (or $t$), 
unlike the term $D(x,t)$.  Thus, it may now be treated as a 
normalization factor that can be essentially ignored, as it 
plays no part in the dynamics or spatial dependence of the 
correlation functions.

The exponential of the right hand side of Eq.\ (\ref{integral}) 
may be rewritten in the form (Spain and Smith 1970),
\begin{equation}
\langle \psi(x) \psi^{\dag}(0) \rangle_{\beta} \simeq 
\frac{i\pi}{\beta v_F \sinh \frac{\pi x}{\beta v_F}} \; . 
\label{corr}
\end{equation}
The above equation again gives us the fermion correlation 
function, except for the phase factor $e^{ik_Fx}$ contributed 
by the chemical potential, which we have ignored by omitting 
the leading phase factors in going from Eq. (\ref{complete_correl}) 
to (\ref{incomp_correl}).  

The commuting of the two fermion operators within the expectation 
changes the divergent factor $D(x,t)$ to $D(-x,-t)$, which merely 
results in a change in sign of $x$ (and $t$) in the integral of 
Eq.\ (\ref{integral}).  From this the relation,
\begin{equation}
\langle \psi^{\dag}(0) \psi(x,t) \rangle_{\beta} = 
\langle \psi(-x,-t) \psi^{\dag}(0) \rangle_{\beta} \; , 
\end{equation}
follows immediately.  This simplifies the derivation of the 
retarded Green's functions, which contain a sum of these terms.

The time dependent correlation functions, as well as all the 
finite temperature Green's functions, may now be calculated.  
To obtain the time dependence of any of these functions we 
simply replace the variable $x$ by $x-v_Ft$.

By introducing a convergence term, $e^{-\alpha q}$, the integral 
may also be solved in closed form as a ratio of gamma functions 
and a term $i/(x+i\alpha)$, which reduce to the form, 
$\pi /\sinh \pi x$, in the limit $\alpha \rightarrow 0$ 
(Emery 1979).  This result uses non-normal ordered terms as 
boson operators which require the introduction of a term 
$\alpha^{-1}$ to retain the fermion field anti-commutation 
relations.  The convergence term introduced in the solution 
of the integral then cancels with the $\alpha^{-1}$ term to 
produce the desired result. 

Preliminary results in obtaining Eq.\ (\ref{coor}) for
a non-interacting spinless model, via bosonization 
has been reported previously by us (Bowen, Gul\'acsi 
and Rosengren 2000). However, Eq.\ (\ref{corr}) was 
known from conformal field theory (Cardy 1984) and it 
is widely recognized (Christe and Henkel 1993) that it is one 
of the most important results of the application of conformal 
invariance to critical phenomena. 

\subsection{The Interacting Spinless Model}

The interaction term for spinless fermions in one dimensional 
is of the form,
\begin{equation}
\hat{H}_{INT} = \frac{\pi}{L}\sum_{q\neq 0} \sum_{k,k'} V(q) 
\hat{c}^{\dag}_{k+q} \hat{c}^{\dag}_{k'-q} 
\hat{c}^{}_{k'} \hat{c}^{}_k \; , 
\end{equation}
the $q=0$ term can be neglected due to the background Coulomb 
interactions with the lattice (Mahan 1981).  In terms of boson 
operators this becomes,
\begin{equation}
\hat{H}_{INT} = \frac{1}{2} \sum_{q>0} q \; V(q) 
\left( \hat{b}^{\dag}_q \hat{b}^{}_q + \hat{b}^{}_q \hat{b}^{\dag}_q 
\right) \; . 
\end{equation}
We commute the second term and renormalize to remove the constant 
interaction energy, then we can express the interaction as,
\begin{equation}
\hat{H}_{INT} = \sum_{q>0} q \, V(q) \hat{b}^{\dag}_q \hat{b}^{}_q \; ,
\end{equation}
which when added to the non-interacting boson Hamiltonian changes 
the dispersion of the bosons, or if $V(q)=V$ is a constant, the 
interaction renormalizes the Fermi velocity,
\begin{equation}
\hat{H}_0 = v_F \sum_q q \; \hat{b}^{\dag}_q \hat{b}^{}_q \qquad 
\longrightarrow \qquad \hat{H}_0 + \hat{H}_{INT} = \sum_{q>0} 
\left(v_F+V(q)\right)q \, \hat{b}^{\dag}_q \hat{b}^{}_q \; . 
\end{equation}

\subsubsection{Finite Temperature Correlation Functions}

The same argument used for the non-interacting correlation 
functions extends to the interacting case.  The only difference 
is in the time dependence of the boson operators $\hat{b}^{}_q$ 
and $\hat{b}^{\dag}_q$, which now have a different dispersion 
$\epsilon_q$ in the time dependence.

The equal time correlation functions are obtained by solving 
the integral,
\begin{equation}
\int_0^{\infty} \frac{2}{q} \left[ \frac{ \cos qx - 1}
{e^{ \beta \epsilon_q} -1} + \frac{e^{iqx}}{2} \right] \; dq \; . 
\end{equation}
where $\epsilon_q$ is the new dispersion function of the bosons.

For $V(q)=V$, the real space interaction is a delta function 
interaction $\delta(x)$, meaning two particles only interact 
if they collide with each other.  For spinless fermions this 
is not possible, as the Pauli exclusion forbids identical 
fermions from being in the same state.  Hence, for a delta 
function potential in this model the result should be the 
same as for a non-interacting system.

In the above calculations we see that the delta function 
interaction changes the finite temperature correlation 
functions by rescaling the Fermi velocity.  This is the 
equivalent of rescaling the inverse temperature, with the 
scaling determined by the coefficient, or strength, of the 
delta function coupling.  Apart from the rescaling, however, 
the distribution function $n(k,\beta)$, is unchanged from the 
non-interacting case.

\subsection{Interacting Spin-$\frac{1}{2}$ Fermions}
\label{spin_case}

Introducing spin to the fermion model fundamentally changes 
the treatment of the interaction terms in terms of boson 
excitations.  Maintaining full $SU(2)$ spin invariance in 
the Hamiltonian requires the use of {\sl non-Abelian bosonization} 
techniques (Witten 1984).

In this section, we shall assume coupling is between only the 
$z$ components of the spin. Hence, we add the $z$ spin component 
to each fermion, which may be in the states denoted by $\uparrow$ 
or $\downarrow$.  The Hamiltonian is then,
\begin{equation}
\hat{H} = v_F \sum_{\sigma} \sum_{k} (k-k_F) 
\hat{c}^{\dag}_{k\sigma} \hat{c}^{}_{k\sigma} + 
\frac{\pi}{L} \sum_{\sigma,\sigma'} \sum_{q\neq 0} 
\sum_{k,k'} V_{\sigma \sigma'}(q) \hat{c}^{\dag}_{k+q,\sigma} 
\hat{c}^{\dag}_{k'-q,\sigma'} \hat{c}^{}_{k'\sigma'} 
\hat{c}^{}_{k\sigma} \; , 
\end{equation}
where $\sigma,\sigma' \in \{ \uparrow , \downarrow \}$.

Writing the Hamiltonian in terms of the boson operators 
$\hat{b}^{}_{q\sigma}$, where $\sigma$ now denotes the spin 
index, and $V_{\parallel}(q)$ and $V_{\perp}(q)$ denote the 
parallel spin and anti-parallel spin components of the 
interaction coupling, we find,
\begin{equation}
\hat{H} = \sum_{\sigma} \sum_{q} \left(v_F + V_{\parallel}(q) \right)
q \; \hat{b}^{\dag}_{q\sigma} \hat{b}^{}_{q\sigma} + 
\sum_{q> 0} q\, V_{\perp}(q) \left( \hat{b}^{\dag}_{q\uparrow} 
\hat{b}^{}_{q\downarrow} + \hat{b}^{\dag}_{q\downarrow} 
\hat{b}^{}_{q\uparrow} \right) \; . 
\end{equation}
The parallel and anti-parallel interaction couplings 
$V_{\parallel}(q)$ and $V_{\perp}(q)$ may not be equal, 
as the parallel component must take into account an exchange 
term due to the indistinguishable nature of the scattering fermions. 
For the anti-parallel case, the scattering fermions are now 
distinguishable and hence no exchange term is present.  
This may lead to the different effective couplings under 
the bosonization of the Hamiltonian (Mahan 1981).

\subsubsection{Spin and Charge Excitations}

Overhauser (1965) has shown that the excitation spectrum of this 
model is completely described by two types of excitations, density 
oscillations and spin waves.  The operators that describe these 
excitations are given by,
\begin{eqnarray}
& \hat{\varrho}_q = \frac{1}{\sqrt{2}} \left( \hat{b}^{}_{q\uparrow} 
+ \hat{b}^{}_{q\downarrow} \right) \; \; \; 
\hat{\varrho}^{\dag}_q 
= \frac{1}{\sqrt{2}} \left( \hat{b}^{\dag}_{q\uparrow} + 
\hat{b}^{\dag}_{q\downarrow} \right) 
\nonumber \\
& \hat{\chi}_q = \frac{1}{\sqrt{2}} \left( \hat{b}^{}_{q\uparrow} 
- \hat{b}^{}_{q\downarrow} \right) \; \; \; 
\hat{\chi}^{\dag}_q 
= \frac{1}{\sqrt{2}} \left( \hat{b}^{\dag}_{q\uparrow} 
- \hat{b}^{\dag}_{q\downarrow} \right) \; , 
\end{eqnarray}
where $\hat{\varrho}_q$ denotes the density excitations, 
and $\hat{\chi}_q$ the spin excitations.  Because the particles 
are presumed to be charged, the density of particles also 
corresponds to the charge density, so we may call the density 
excitations charge excitations.  The commutators between 
different operators all vanish, meaning the spin and charge 
excitations are independent of each other.

The commutators of the charge and spin operators with the 
Hamiltonian give the results,
\begin{eqnarray}
& [ \hat{H} , \hat{\varrho}_q ] 
= -\left(v_F + V_{\parallel}(q)+ V_{\perp}(q) \right) q \; 
\hat{\varrho}_q = -v_{\varrho}(q)q \, \hat{\varrho}_q 
\nonumber \\
& [ \hat{H} , \hat{\chi}_q ] 
= -\left(v_F + V_{\parallel}(q) - V_{\perp}(q) \right) q \; 
\hat{\chi}_q = -v_{\chi}(q) q \,\hat{\chi}_q \; , 
\end{eqnarray}
which again entails we can write the Hamiltonian in the form,
\begin{equation}
\hat{H} = \sum_q v_{\varrho}(q)q\, \hat{\varrho}^{\dag}_q 
\hat{\varrho}^{}_q + \sum_q v_{\chi}(q) q \; 
\hat{\chi}^{\dag}_q \hat{\chi}^{}_q + \hat{C} \; , 
\end{equation}
the two types of excitations thus exist independently of 
each other, and may obey different dispersion laws.  Any 
difference in the dispersion means that the two types of 
excitation may separate with time, leading to a phenomena 
known as {\sl spin-charge separation}.

The operator $\hat{C}$ contains the $q=0$ components which 
account for the fermion number operators which may also be 
split into spin and charge operators.  Again, this term 
relates to the ground state energies of the boson states, 
and represents the chemical potential of the fermion system.

This can be achieved by establishing charge and spin ground 
states,
\begin{equation}
\hat{N}_{\varrho} = \frac{1}{2}(\hat{N}_{\uparrow} + 
\hat{N}_{\downarrow}) \qquad \text{and} \qquad \hat{N}_{\chi} 
= \frac{1}{2}(\hat{N}_{\uparrow} - \hat{N}_{\downarrow}) \; , 
\end{equation}
respectively.  These operators may then be separated into the 
respective spin and charge components of the Hamiltonians 
(Voit 1995). 

\subsubsection{Bosonization in Terms of Spin and Charge Excitations}
\label{spin_charge}

Because the spin and charge operators are constructed from the 
boson operators $\hat{b}^{}_{q\sigma}$, we can rewrite the 
bosonization identity between fermions and bosons in terms of 
the spin and charge operators.  This allows us to easily separate 
the expectation values into a product of spin and charge components.

We note that, for $s= 1, -1$ where $\sigma=\uparrow, \downarrow$, 
respectively, we have,
\begin{equation}
\hat{b}^{}_{q\sigma} = \frac{1}{\sqrt{2}} 
\left( \hat{\varrho}^{}_q + s\hat{\chi}_q \right) \; , \; \; \; 
\hat{b}^{\dag}_{q\sigma} = \frac{1}{\sqrt{2}} 
\left( \hat{\varrho}^{\dag}_q + s\hat{\chi}^{\dag}_q \right) \; 
\end{equation}
and hence,
\begin{eqnarray}
& \varphi^{}_{\sigma}(x) 
= -\sum_q \sqrt{\frac{\pi}{Lq}} e^{iqx} 
\left( \hat{\varrho}^{}_q + s \hat{\chi}^{}_q \right) 
= \varphi^{}_{\varrho}(x) + s \varphi^{}_{\chi}(x) 
\nonumber \\
& \varphi^{\dag}_{\sigma}(x) 
= -\sum_q \sqrt{\frac{\pi}{Lq}} e^{-iqx} 
\left( \hat{\varrho}^{\dag}_q + s \hat{\chi}^{\dag}_q \right)
= \varphi^{\dag}_{\varrho}(x) + s \varphi^{\dag}_{\chi}(x) \; . 
\end{eqnarray}
As the spin and charge operators commute, we can split any 
exponentials of the sum of these operators into a product of 
exponentials of each operator.  These exponentials will then 
commute past each other, and the single fermion operators may 
be represented as,
\begin{equation}
\hat{\psi}_{\sigma}(x) = \hat{F}_{\sigma} \lambda_{\sigma}(x) 
e^{\varphi^{\dag}_{\varrho}(x)} e^{-\varphi_{\varrho}(x)}  
e^{s\varphi^{\dag}_{\chi}(x)} e^{-s\varphi_{\chi}(x)} \; . 
\label{spin_char_fermi}
\end{equation}

The expectation values of the fermion operators then become 
a product of two boson expectation values, with the charge 
operators traced over a set of charge excitation states, the 
spin operators traced over a set of spin excitation states.  
The boson excitation states can be factored into separate 
spin and charge number states, as these excitations are 
independent.

For single particle correlation functions involving fermions 
of opposite spin, the Klein factors will no longer vanish.  
The correlations of the most relevance in these models, such 
as singlet and triplet superconductive pairing, and magnetic 
susceptibility, all involve two particle correlation functions 
in which the Klein factors will again cancel.

Because the spin-charge boson Hamiltonian is non-interacting 
even when the underlying fermions are interacting, the 
expectation values obtained are identical to the non-interacting 
fermion case, with only the factor $\sqrt{2\pi/Lq}$ changed to 
$\sqrt{\pi/Lq}$.  The dependence on the spin state $\sigma$ 
given by $s=\pm 1$ vanishes in the correlation functions, as 
only factors of $s^2=1$ appear.  In taking the continuum limit, 
where the sums in the exponentials become integrals, the factor 
of $L/2\pi$ does not cancel the $\pi /L$, and each correlation 
retains an extra power of $1/2$.

In the case of a real space delta function interaction, we use 
the Pauli exclusion principle to set $V_{\parallel} = 0$, and 
we have only a delta function interaction $V=V_{\perp}$, for 
fermions of opposite spin. The time dependent correlation 
functions may then be written explicitly as,
\begin{equation}
\langle \psi_{\sigma}(x,t) \psi^{\dag}_{\sigma}(0,0) \rangle_{\beta} 
=  [ \pi / \beta v_{\varrho} \sinh 
\frac{\pi (x-v_{\varrho}t)}{\beta v_{\varrho}} ]^{1/2} \; 
[ \pi / \beta v_{\chi}\sinh 
\frac{\pi (x-v_{\chi}t)}{\beta v_{\chi}} ]^{1/2} \; , 
\end{equation}
where $v_{\varrho} = v_F + V$ and $v_{\chi} = v_F - V$.

\section{The Tomonaga-Luttinger Model}

The Tomonaga-Luttinger model is an exactly soluble model of 
a one dimensional interacting electron gas.  Tomonaga (1950)
developed a model of interacting fermions 
in one dimension in which the density operators had 
approximately bosonic commutation relations.  This model 
contains fermions with linear dispersion and a finite 
bandwidth.

In order to make the bosonic commutation relations of the 
density operators exact, Luttinger (1960) extended 
the model to contain two independent branches of fermions with 
infinite linear dispersion.  One branch has a spectrum given 
by $\epsilon_k = kv_F$ and the other by $\epsilon_k = -kv_F$. The 
two branches are called {\sl left movers} and {\sl right movers}, 
respectively. 

Interactions are restricted to small momentum transfer forward 
scattering processes in order to make an exact solution possible.  
The effects of adding large momentum transfer interactions to the 
Tomonaga-Luttinger model is the subject of the next chapter.

The model was correctly solved by Mattis and Lieb (1965)
by diagonalisation of the Hamiltonian. The zero temperature 
single particle Green's functions were then calculated explicitly 
by Theumann (1967), and Dover (1968). The representation of the 
fermion operators in terms of boson operators, discovered 
independently by both Luther and Peschel (1974), and Mattis
(1974), made further calculation of correlation functions 
straight forward.  The single particle spectral functions 
were calculated independently by Meden and Sch\"{o}nhammer 
(1992) and Voit (1993). 

The importance of the Tomonaga-Luttinger model lies with the 
conjecture by Haldane (1981) that the model is the asymptotically 
exact model for gap-less one dimensional interacting fermions.  
The essential physics of each of the gap-less degrees of 
freedom is then only dependent on the values of two parameters, 
which may be determined by methods such as perturbative 
renormalization or numerical solutions.

In this chapter we review the bosonization solution for the 
Tomonaga-Luttinger model.  We obtain the finite temperature 
single particle correlation functions.  Higher order correlation 
functions and Green's functions may be determined by similar 
methods, and their zero temperature form is covered elsewhere
(Mattis and Lieb 1965; Voit 1995). The calculation of the retarded 
single particle Green's function at finite temperature for 
a delta function interaction is also contained in the article 
by Emery (1979) as well as some finite temperature 
susceptibilities.

\subsection{The Luttinger Hamiltonian}

The model contains two species of fermions, left movers and 
right movers, that are represented by the fermion operators 
$\hat{c}^{}_{k\sigma}$, $\hat{d}^{}_{k\sigma}$, and their adjoints.
The Hamiltonian consists of the kinetic term,
\begin{equation}
\hat{H}_0 = v_F \sum_{k,\sigma} (k-k_F) \; 
\hat{c}^{\dag}_{k,\sigma} \hat{c}^{}_{k,\sigma} + v\sum_{k,\sigma}
(-k-k_F) \; \hat{d}^{\dag}_{k,\sigma} \hat{d}^{}_{k,\sigma} \; , 
\end{equation}
and interaction terms,
\begin{eqnarray}
& \hat{H}_{INT} 
= \frac{\pi}{L} \sum_{\sigma, \sigma'} \sum_q g_{2,\sigma \sigma'} 
\hat{\rho}^{}_{c,\sigma}(q) \hat{\rho}^{}_{d,\sigma'}(-q) 
\nonumber \\
&+ \frac{2\pi}{L} \sum_{\sigma, \sigma'} \sum_{q>0} 
g_{4,\sigma \sigma'} \left[ \hat{\rho}^{}_{c,\sigma}(q) 
\hat{\rho}^{}_{c,\sigma'}(-q) + \hat{\rho}^{}_{d,\sigma}(-q)
\hat{\rho}^{}_{d,\sigma'}(q) \right] \; . 
\label{new}
\end{eqnarray}

The interaction Hamiltonian contains only forward scattering 
terms, backward and Umklapp scattering are not included in 
the Tomonaga-Luttinger model.  The coupling coefficients are 
labelled according to the $g$-ology prescription. The model 
is a generalization of the Tomonaga-Luttinger model, and 
represents the continuum limit of possible lattice models.  
The kinetic term is equivalent to that of the Luttinger model, 
with two linear branches of fermions. 

The interactions consist of four terms, with coupling constants 
labelled by $g_i$, for $i=1,2,3,4$.  Each coupling constant may 
also be spin dependent $g_{i \sigma \sigma'} = g_{i\parallel} + 
g_{i\perp}$, where $\parallel$ and $\perp$, refer to parallel 
and anti-parallel spins, respectively.

The terms $g_2$ and $g_4$ correspond to forward scattering, or 
small momentum transfer scattering.  The $g_4$ term 
corresponds to two particles on the same branch, and $g_2$ to one 
particle on each branch scattering. 

The backward scattering term $g_1$ relates to large momentum 
transfer of particles onto opposite branches. In models 
including spin, the parallel component of the backward 
scattering term may be viewed as an effective forward scattering 
term, as the interaction represents the exchange term of the $g_2$ 
term. In this case the $g_{2\parallel}$ term may be rewritten to 
include the $g_{1\parallel}$ contribution.  For anti-parallel 
spins the scattering electrons are distinguishable, and hence 
do not have an exchange term. 

The Umklapp term $g_3$ is generally only significant in lattice 
models when the band filling is commensurate with the lattice, 
that is $4k_F$ is equal to a reciprocal lattice vector.  Two 
particles on the same branch may then scatter onto the opposite 
branch, and this term becomes important.

The density operators in Eq.\ (\ref{new}) have been defined as,
\begin{equation}
\hat{\rho}^{}_{c,\sigma}(q) = \sum_k \hat{c}_{k+q,\sigma}^{\dag}
\hat{c}^{}_{k,\sigma} \; , \; \; \; 
\hat{\rho}^{}_{d,\sigma}(q) = \sum_k \hat{d}_{k+q,\sigma}^{\dag}
\hat{d}^{}_{k,\sigma} \; . 
\end{equation}

The commutators of the density operators are obtained from,
\begin{equation}
[ \hat{\rho}^{}_{c,\sigma}(q), \hat{\rho}^{}_{c,\sigma'}(-q') ] = 
\delta_{\sigma \sigma'} \sum_{k} \left( \hat{c}_{k+q,\sigma}^{\dag} 
\hat{c}^{}_{k+q',\sigma} - \hat{c}_{k+q-q',\sigma}^{\dag} 
\hat{c}^{}_{k,\sigma} \right) \; . 
\end{equation}
Since the operators must be normal ordered in order to subtract 
them properly, we find the commutator comes out as,
\begin{equation}
[ \hat{\rho}^{}_{c,\sigma'}(-q') , \hat{\rho}^{}_{c,\sigma}(q) ] = 
\delta_{\sigma \sigma'} \delta_{q,-q'} \frac{Lq}{2\pi} \; , 
\end{equation}
which gives an approximately bosonic commutation relation.

Thus, we can define boson operators to make the commutation 
relations exact. Even though there are now two species of 
fermions, we can in fact, assign a single species of boson 
operator for all $q$ (Emery 1979).  This is done by assigning, 
for $q>0$,
\begin{equation}
\hat{b}^{}_{q\sigma} = \sqrt{\frac{2\pi}{Lq}} 
\hat{\rho}^{}_{c,\sigma}(-q) \; , \; \; \; 
\hat{b}^{\dag}_{q\sigma} = \sqrt{\frac{2\pi}{Lq}}
\hat{\rho}^{}_{c,\sigma}(q) \; , 
\end{equation}
and similarly,
\begin{equation}
\hat{b}^{}_{-q\sigma} = \sqrt{\frac{2\pi}{Lq}}
\hat{\rho}^{}_{d,\sigma}(q) \; , \; \; \; 
\hat{b}^{\dag}_{-q\sigma} = \sqrt{\frac{2\pi}{Lq}}
\hat{\rho}^{}_{d,\sigma}(-q) \; . 
\end{equation}
These operators now obey exact boson commutation relations, 
$[ \hat{b}^{}_{q\sigma}, \hat{b}^{\dag}_{q'\sigma'} ] = 
\delta_{qq'} \delta_{\sigma \sigma'}$.

All other boson fields generated as functions of 
$\hat{b}^{}_{q\sigma}$ and $\hat{b}^{\dag}_{q\sigma}$, in 
previous chapters may now be defined analogously in terms 
of these operators.  It should be noted, however, that the 
boson operators for $q>0$ correspond to the right moving 
fermions, and the $q<0$ bosons to the left moving fermions, 
so the boson field operators corresponding to the left and 
right movers must have sums over the appropriate range of 
momenta $q$.

The commutator of the kinetic term of the Hamiltonian with 
any boson operator gives,
\begin{equation}
[ \hat{H}_0 ,\hat{b}^{}_{q\sigma} ] = - v_F |q| 
\hat{b}^{}_{q\sigma} = -\epsilon(q) \hat{b}^{}_{q\sigma} \; , 
\end{equation}
where the dispersion is $\epsilon(q)$ is $v_Fq$ for $q>0$, 
and $-v_Fq$ for $q<0$.

From Kr\"{o}nig's relation, the kinetic term can be written 
in a quadratic form of the density operators,
\begin{equation}
\hat{H}_0 = \sum_{\sigma} \sum_{q>0} v_Fq \left( 
\hat{b}^{\dag}_{q\sigma} \hat{b}^{}_{q\sigma} + 
\hat{b}^{\dag}_{-q\sigma} \hat{b}^{}_{-q\sigma} \right) \; . 
\end{equation}
Also, because the bosons are defined as excitations above 
an $N$ particle ground state, the Hamiltonian must include 
terms that include the energy of the different bosonic ground 
states.  These terms are not required for the calculations in 
this chapter, and are hence omitted.

The interaction term in terms of the boson operators is obtained 
by direct substitution,
\begin{eqnarray}
\hat{H}_{INT} 
&= \sum_{\sigma, \sigma'} \sum_{q>0} g_{2,\sigma \sigma'} q 
\left( \hat{b}^{\dag}_{q\sigma} \hat{b}^{\dag}_{-q\sigma'} + 
\hat{b}^{}_{q\sigma} \hat{b}^{}_{-q\sigma'} \right) 
\nonumber \\
&+ \sum_{\sigma, \sigma'} \sum_{q>0} g_{4,\sigma \sigma'} q 
\left( \hat{b}^{\dag}_{q\sigma} \hat{b}^{}_{q\sigma'} +
\hat{b}^{\dag}_{-q\sigma} \hat{b}^{}_{-q\sigma'} \right) \; . 
\end{eqnarray}

\subsection{Diagonalising the Hamiltonian}

The eigenvalues and eigenstates of the Luttinger model were 
first solved by diagonalisation of the spinless Hamiltonian 
by Mattis and Lieb (1965).  In this section we diagonalise 
the bosonic Hamiltonian by a Bogolubov transformation 
(Emery 1979, Mahan 1981, Voit 1995). 

For each degree of freedom in the model, we obtain a 
renormalized velocity parameter, $v$, and a renormalized 
coupling parameter, $K$, that completely describe the 
properties of the system.

\subsubsection{Diagonalisation for Spinless Fermions}

We start by diagonalising the case with spinless fermions, 
as this generalizes easily to the case with spin. The boson 
Hamiltonian in this case becomes,
\begin{equation}
\hat{H} = \sum_{q>0} (v_F+g_4(q)) q 
\left( \hat{b}^{\dag}_{q} \hat{b}^{}_{q} + 
\hat{b}^{\dag}_{-q} \hat{b}^{}_{-q} \right) + 
\sum_{q>0} g_{2}(q)q \left( \hat{b}^{\dag}_{q} \hat{b}^{\dag}_{-q} 
+ \hat{b}^{}_{q} \hat{b}^{}_{-q} \right) \; . 
\label{luttinger_form}
\end{equation}
We define the operator,
\begin{equation}
\hat{S} = i \sum_{q>0} \phi_q \left( \hat{b}^{\dag}_q 
\hat{b}^{\dag}_{-q} + \hat{b}^{}_q \hat{b}^{}_{-q} \right) \; , 
\end{equation}
which has the commutation relations $[ \hat{S},\hat{b}^{\dag}_{q} ] 
= -i \phi_q \hat{b}^{}_{-q}$ and $[ \hat{S},\hat{b}_{-q} ] = -i 
\phi_q \hat{b}^{\dag}_{q}$, with $\phi_q = \phi_{-q}$.  The boson 
operators then transform to give,
\begin{eqnarray}
e^{i\hat{S}} \hat{b}^{\dag}_{q} e^{-i\hat{S}} 
&= \hat{b}^{\dag}_{q} \cosh \phi_q + \hat{b}^{}_{-q} \sinh \phi_q 
\nonumber \\
e^{i\hat{S}} \hat{b}^{}_{-q} e^{-i\hat{S}} 
&= \hat{b}^{\dag}_{q} \sinh \phi_q + \hat{b}^{}_{-q} \cosh \phi_q \; , 
\end{eqnarray}
and the Hamiltonian may be written in diagonal form,
\begin{equation}
\hat{H}_D = e^{i\hat{S}} \hat{H} e^{-i\hat{S}} = 
\sum_{q} v_F |q| \left[ \left( 1 + \frac{g_4(q)}{v_F} \right)^2 
- \left( \frac{g_2(q)}{v_F} \right)^2 \right]^{1/2}  
\hat{b}^{\dag}_q \hat{b}^{}_q \; , 
\end{equation}
when the condition,
\begin{equation}
\tanh 2\phi_q = \frac{g_2(q)}{v_F + g_4(q)}
\end{equation}
is satisfied.  This condition may be rewritten in terms of an 
effective coupling constant $K(q)$, which appears in correlation 
functions, by using the identity, $\tanh^{-1} x = \frac{1}{2} 
\ln ( 1+x) - \frac{1}{2} \ln (1-x)$, to obtain,
\begin{equation}
K(q) = e^{2\phi_q} = \left( \frac{v_F + g_4(q) - g_2(q)}
{v_F + g_4(q) + g_2(q)} \right)^{\frac{1}{2}} \; . 
\end{equation}

\subsubsection{The Spin-$\frac{1}{2}$ Hamiltonian}
\label{spin_diag}

The addition of spin into the Hamiltonian does not change 
from the spinless case if the spins of the bosons are identical, 
apart from the addition of a sum over spin states.  The terms 
that involve boson operators of opposite spins add terms equivalent 
to those introduced in the single branch model considered in 
Section \ref{spin_case}.

As done previously, we may obtain the form of the Hamiltonian 
we need, by transforming to the spin and charge variables,
\begin{eqnarray}
& \hat{\varrho}^{}_{q} = \frac{1}{\sqrt{2}} 
\left( \hat{b}^{}_{q\uparrow} + \hat{b}^{}_{q\downarrow} \right) 
\; , \; \; \; 
\hat{\varrho}^{\dag}_{q} = \frac{1}{\sqrt{2}} 
\left( \hat{b}^{\dag}_{q\uparrow} + 
\hat{b}^{\dag}_{q\downarrow} \right) 
\nonumber \\
& \hat{\chi}^{}_{q} = \frac{1}{\sqrt{2}} 
\left( \hat{b}^{}_{q\uparrow} - \hat{b}^{}_{q\downarrow} \right) 
\; , \; \; \; 
\hat{\chi}^{\dag}_{q} = \frac{1}{\sqrt{2}} 
\left( \hat{b}^{\dag}_{q\uparrow} - \hat{b}^{\dag}_{q\downarrow} 
\right) \; . 
\end{eqnarray}
Under such a change in variables the interaction couplings 
transform as,
\begin{eqnarray}
& g_{2\varrho} = \frac{1}{2} 
\left( g_{2\parallel} + g_{2\perp} \right) 
\; , \; \; \; 
g_{4\varrho} = \frac{1}{2} 
\left( g_{4\parallel} + g_{4\perp} \right) 
\nonumber \\
& g_{2\chi} = \frac{1}{2} 
\left( g_{2\parallel} - g_{2\perp} \right) 
\; , \; \; \; 
g_{4\chi} = \frac{1}{2} 
\left( g_{4\parallel} - g_{4\perp} \right) \; . 
\end{eqnarray}
The transformation then gives a total Hamiltonian consisting of 
the sum of two Hamiltonians of the spinless Luttinger form, as 
in Eq.\ (\ref{luttinger_form}), one for the spin and one for 
the charge terms.  These are independent and can be diagonalised 
by using the sum of the operators,
\begin{equation}
\hat{S}_{\varrho} = i \sum_{q>0} \phi_{\varrho}(q) 
\left( \hat{\varrho}^{\dag}_q \hat{\varrho}^{\dag}_{-q} + 
\hat{\varrho}^{}_q \hat{\varrho}^{}_{-q} \right) \; , \; \; \; 
\hat{S}_{\chi} = i \sum_{q>0} \phi_{\chi}(q) 
\left( \hat{\chi}^{\dag}_q \hat{\chi}^{\dag}_{-q} + 
\hat{\chi}^{}_q \hat{\chi}^{}_{-q} \right) \; . 
\label{spin_diag_op}
\end{equation}
The diagonal Hamiltonian has spin and charge terms with the 
renormalized charge and spin velocities,
\begin{equation}
v_{\varrho}(q) = \left( (v_F +g_{4\varrho}(q))^2 - 
g_{2\varrho}^2(q) \right)^{1/2} \; , \; \; \; 
v_{\chi}(q) = \left( (v_F +g_{4\chi}(q))^2 - 
g_{2\chi}^2(q) \right)^{1/2} \; , 
\end{equation}
and the renormalized effective coupling constants become,
\begin{equation}
K_{\varrho}(q) = \left( \frac{v_F + g_{4\varrho}(q) - 
g_{2\varrho}(q)}{v_F + g_{4\varrho}(q) + g_{2\varrho}(q)} 
\right)^{1/2} \; , \; \; \; 
K_{\chi}(q) = \left( \frac{v_F + g_{4\chi}(q) - g_{2\chi}(q)}
{v_F + g_{4\chi}(q) + g_{2\chi}(q)} \right)^{1/2} \; . 
\label{spin_K}
\end{equation}

\subsection{Finite Temperature Correlation Functions}

To obtain the single particle correlation functions, we must 
again express the fermion operators in terms of the diagonalised 
boson operators.  The expectation over the fermion operators may 
be rewritten in the form,
\begin{eqnarray}
\langle \hat{\psi}^{}_{\sigma}(x,t) 
\hat{\psi}^{\dag}_{\sigma'}(0,0) \rangle_{\beta} 
&= \frac{1}{\cal{Z}} {\rm{Tr}} \; 
e^{i\hat{S}} e^{-\beta \hat{H}} e^{-i\hat{S}} 
e^{i\hat{S}} \hat{\psi}^{}_{\sigma}(x,t) 
\hat{\psi}^{\dag}_{\sigma'}(0,0) e^{-i\hat{S}} 
\nonumber \\
&= \frac{1}{\cal{Z}} {\rm{Tr}} \; 
e^{-\beta \hat{H}_D} e^{i\hat{S}} 
\hat{\psi}^{}_{\sigma}(x,t) \hat{\psi}^{\dag}_{\sigma'}(0,0) 
e^{-i\hat{S}} \; . 
\end{eqnarray}
As the operator $e^{i\hat{S}}$ is unitary, it can be inserted 
between operators in the form $1=e^{-i\hat{S}}e^{i\hat{S}}$.  
Also, the trace is invariant under cyclic permutations of the 
operators.  The non-diagonal and diagonal Hamiltonians are 
denoted by $\hat{H}$ and $\hat{H}_D$, respectively.

\subsubsection{The Spinless Case}

The correlation functions can be determined by replacing the 
transformed fermion operators with the transformed boson field 
operators.  Calculating this for the right moving fermion field 
$\psi_R (x)$, which is represented by bosons with $q>0$, we have,
\begin{eqnarray}
& \left\langle e^{i\hat{S}} \hat{\psi}^{}_{R}(x) 
\hat{\psi}^{\dag}_{R}(0) e^{-i\hat{S}} \right\rangle \simeq D(x,0) 
\Bigg\langle e^{i\hat{S}} \exp \left( \sum_{q>0} 
\sqrt{\frac{2\pi}{Lq}} (1-e^{-iqx}) \hat{b}^{\dag}_q \right) 
e^{-i\hat{S}} 
\nonumber \\
& \times e^{i\hat{S}} \exp \left( \sum_{q>0} \sqrt{\frac{2\pi}{Lq}} 
(e^{iqx}-1) \hat{b}_q \right) e^{-i\hat{S}} \Bigg\rangle \; , 
\end{eqnarray}
where $D(x,0)$ is the divergent term generated by ordering the 
boson fields.  The transformation leads to terms involving both 
left and right moving bosons, represented as $q>0$ and $q<0$ 
terms.  These are independent and may be separated, along with 
the diagonalised Hamiltonian, into the product of expectation 
values for left and right moving fields.  The right moving fields 
obtain a $\cosh \phi_q$ dependence,
\begin{equation}
\left\langle \exp \left( \sum_{q>0} \sqrt{\frac{2\pi}{Lq}} 
(1-e^{-iqx}) \hat{b}^{\dag}_q \cosh \phi_q \right) \exp 
\left( \sum_{q>0} \sqrt{\frac{2\pi}{Lq}} (e^{iqx}-1) 
\hat{b}^{}_q \cosh \phi_q \right) \right\rangle_{R} \; , 
\end{equation}
and the left movers a $\sinh \phi_q$ term,
\begin{equation}
\left\langle \exp \left( \sum_{q>0} \sqrt{\frac{2\pi}{Lq}} 
(1-e^{-iqx}) \hat{b}^{}_{-q} \sinh \phi_q \right) \exp 
\left( \sum_{q>0} \sqrt{\frac{2\pi}{Lq}} (e^{iqx}-1) 
\hat{b}^{\dag}_{-q} \sinh \phi_q \right) \right\rangle_{L} \; .  
\end{equation}

If the coupling constants are considered independent of the 
momentum $q$, which makes $\phi_q = \phi$, a constant, and 
the $\sinh \phi$ and $\cosh \phi$ terms can be brought outside 
the summation.  The form of these correlation functions does 
not change except for the introduction of an extra power term, 
$1+\gamma = \cosh^2 \phi$ or $\gamma=\sinh^2 \phi$, respectively.  
The term $\gamma$ depends on the value of the coupling constant 
$K$, and is given by,
\begin{equation}
\gamma = \frac{1}{4}\left( K+\frac{1}{K}-2\right) \; . 
\end{equation}

As can be seen from the definitions, the mixing of fermion 
species and the anomalous power law scalings occur only in 
the case of scattering between fermions on different branches 
of the dispersion, $g_2 \neq 0$, which implies $K\neq 1$.

With $\phi$ a constant, the divergent term $D(x,0)$ may be 
split into two using the identity, $\cosh^2 \phi - \sinh^2 \phi =1$, 
to add a term $\pm e^{iqx}$ to the right and left moving components, 
respectively.  The right moving term is equivalent to the term given 
in Eq.\ (\ref{integral}), except for a constant $1 + \gamma$, 
which becomes an exponent.

The left moving term contains the operators in a reverse order 
compared to the right moving term.  This produces an additional 
term, $2\cos qx -2$, which removes the, $e^{iqx}$, component 
added by splitting the, $D(x,0)$, term and leaves an $x$ 
dependent term $e^{-iqx}$ and a constant.  Rearranging, we find,
\begin{equation}
\int_0^{\infty} \frac{2}{q} \left[ \frac{\cos qx - 1}
{e^{\beta v_F q} - 1} + \frac{e^{-iqx}}{2} \right] dq = 
\ln \left[ \Gamma\left( \frac{ix}{\beta v_F} \right) 
\Gamma\left( 1-\frac{ix}{\beta v_F} \right) \right] + Q' \; , 
\label{integral2}
\end{equation}
with a new divergent constant $Q'$, given by,
\begin{equation}
Q' = \int_0^{\infty} \frac{e^{-\beta v_F q}-2}{q} \; dq \; . 
\end{equation}

Ignoring the divergent normalization terms $Q$ and $Q'$, and 
the phase factor, the equal time correlation function may now 
be written,
\begin{equation}
\langle \psi_R(x) \psi_R^{\dag}(0) \rangle_{\beta} \simeq 
\left( \frac{i\pi}{\beta v_F \sinh \frac{\pi x}{\beta v_F}} 
\right)^{\alpha} \; , 
\end{equation}
where, $\alpha = 1+2\gamma$.  This reduces to, $\alpha = 1$, 
in the case $K=1$.  The correlation function for left movers 
$\psi_L(x)$, is derived analogously, and gives the same result, 
at least for equal time correlations. This result agrees with 
the conformal field theory results for finite temperature
(Frahm and Korepin 1990). 

The inclusion of time dependence into the above results 
requires slightly more care, as the terms are obtained 
from a combination of left and right moving fields.  
The right moving fields maintain the $x-v_Ft$ dependence 
on time, whereas the left moving fields have an $x+v_Ft$ 
dependence.

The two properties that determine the non-Fermi liquid 
behaviour in one dimension are spin-charge separation 
and non-universal power law behaviour (Metzner and Di 
Castro 1993; Gul\'acsi 1997; Voit 1998). We have shown 
that these two properties are valid also at finite 
temperature. Previously we proved that spin-charge 
separation is determined only by scattering at the 
same Fermi point.  While the above results show that 
the anomalous power law behaviour occurs due to 
scattering across the two different Fermi points. 

\subsubsection{The Case With Spin}

To calculate the correlation functions including spin in 
the model, we must first transform into the spin and charge 
variables, so as to make the Hamiltonian diagonalizable in 
these variables, as shown in Section \ref{spin_diag}.

The correlation functions may be expressed as a product of 
the spin and charge components, where each has the same 
form as the spinless case presented above.  For example,
\begin{eqnarray}
& \left\langle e^{i\hat{S}} \hat{\psi}^{}_{R}(x) 
\hat{\psi}^{\dag}_{R}(0) e^{-i\hat{S}} \right\rangle \simeq D(x,0) 
\nonumber \\ 
& \times \prod_{\mu = \varrho , \chi} \bigg\langle e^{i\hat{S}_{\mu}} 
\exp \left( \sum_{q>0} \sqrt{\frac{\pi}{Lq}} (1-e^{-iqx}) 
\hat{\mu}^{\dag}_q \right) e^{-i\hat{S}_{\mu}} 
\nonumber \\
& \times e^{i\hat{S}_{\mu}} \exp \left( \sum_{q>0} 
\sqrt{\frac{\pi}{Lq}} (e^{iqx}-1) \hat{\mu}^{}_q \right) 
e^{-i\hat{S}_{\mu}} \bigg\rangle_{\mu} \; . 
\end{eqnarray}

Under the transformation by the diagonalisation operators, 
$\hat{S}$, the correlation functions again split into 
independent left and right moving components.  Each of 
these components is equivalent to a spinless correlation 
function, with the exception that each correlation exponent 
is now multiplied by a half.  The charge and spin components 
also now have separate correlation exponents $\gamma_{\varrho}$ 
and $\gamma_{\chi}$,
\begin{equation}
\gamma_{\varrho} = \frac{1}{8}\left( K_{\varrho} + 
\frac{1}{K_{\varrho}}-2\right) \; , \; \; \; 
\gamma_{\chi} 
= \frac{1}{8}\left( K_{\chi} + \frac{1}{K_{\chi}} - 2 \right) \; , 
\end{equation}
and we obtain the result,
\begin{equation}
\left\langle \hat{\psi}_{R}(x) \hat{\psi}^{\dag}_{R}(0) 
\right\rangle \simeq
e^{ik_Fx} \prod_{\mu = \varrho , \chi} 
\frac{\pi}{\sqrt{\beta v_{\mu} \sinh \frac{\pi x}{\beta v_{\mu}}}} 
\left( \frac{\pi}{\beta v_{\mu} \sinh \frac{\pi x}{\beta v_{\mu}}} 
\right)^{\gamma_{\mu}} \; , 
\end{equation}
for the right movers.

The time dependence of the correlation functions may be 
obtained by substituting $x+v_{\mu}t$ and $x-v_{\mu}t$ in 
each left and right moving term, respectively.  Thus the 
full time dependent correlation functions are given by,
\begin{equation}
\left\langle \hat{\psi}_{R}(x) \hat{\psi}^{\dag}_{R}(0) 
\right\rangle \simeq
e^{ik_Fx} \prod_{\mu = \varrho , \chi} 
\left( \frac{\pi}{\beta v_{\mu} \sinh 
\frac{\pi (x+v_{\mu}t)}{\beta v_{\mu}}} 
\right)^{\frac{\gamma_{\mu}}{2}} 
\left( \frac{\pi}{\beta v_{\mu} \sinh 
\frac{\pi (x-v_{\mu}t)}{\beta v_{\mu}}} 
\right)^{\frac{1 + \gamma_{\mu}}{2}} \; , 
\end{equation}
and similarly for the left movers.

\section{The Hubbard Model}

The Hubbard model was constructed in an attempt to model the 
correlations of electrons occupying the $d$ band of transition 
metals (Gutzwiller 1963, Hubbard 1963). In contrast to the free 
electron model of the conduction band electrons, the Hubbard 
model is a simple {\sl tight binding} model. In a tight binding 
model each electron is considered as localised on a given atomic 
site, and occasionally the electron will hop to another atomic 
site, normally a nearest neighbour.

The simplest model assumes each site consists of a single 
$s$-band orbital, and may be occupied by, at most, one spin 
up electron and one spin down electron.  The interactions 
between the electrons are strongest when they occupy the 
same atomic site, and all other inter-site interactions 
are ignored. The extended Hubbard model includes a contribution 
from nearest neighbour interactions between electrons.

As presented in Section \ref{intro}, an exact solution for 
the Hubbard model been found, where the solution depends on 
an extension of the Bethe ansatz.  The ground state energy, 
wave function, and chemical potential for the one dimensional 
Hubbard model were first found by Lieb and Wu (1968).  
However, the wavefunction is far too complex to obtain 
expectation values and correlation functions by current methods.  
Takahashi (1972) later formulated the thermodynamic Bethe Ansatz 
equations for finite temperature, from which thermodynamics 
properties of the model may be derived.

Despite the simple nature of the model the system shows 
surprising complexity of behaviour.  
The Hubbard model has also been used extensively in the 
examination of the possible failure of Fermi liquid theory 
in dimensions higher that one.  The higher dimensional models 
are not soluble by the same methods as the one dimensional 
model, and techniques such as higher dimensional bosonization 
are yet to prove useful.

\subsection{The Hubbard Hamiltonian}

The model has a Hamiltonian of the form,
\begin{equation}
\hat{H} = -\sum_{\sigma}\sum_{j,k} \tau_{jk} 
\left( \hat{a}^{\dag}_{j,\sigma} \hat{a}^{}_{k,\sigma} + 
\hat{a}^{\dag}_{k,\sigma} \hat{a}^{}_{j,\sigma} \right) + 
\mu \sum_{\sigma} \sum_j \hat{n}_{j, \sigma} + 
U \sum_{\sigma}\sum_j \hat{n}_{j, \uparrow} 
\hat{n}_{j, \downarrow} \; , 
\end{equation}
where the operators $\hat{a}^{\dag}_{j,\sigma}$ and 
$\hat{a}^{}_{j,\sigma}$ are the creation and annihilation operators, 
respectively, for an electron in a Wannier state on the atomic 
lattice site $j$ with spin $\sigma = \{ \uparrow , \downarrow \}$.

The first term is the kinetic term, where the numbers $\tau_{jk}$ 
are the transition probabilities for an electron to hop from site 
$j$ to site $k$.  To further simplify the model, the inter-atomic 
hopping can be restricted to nearest neighbour sites, where we 
write $\tau_{jk} = \tau$ for $j=k\pm 1$, and $\tau_{jk}=0$ 
otherwise.

The second term contains the chemical potential $\mu$, and 
the number operator $\hat{n}_{j, \sigma} = 
\hat{a}^{\dag}_{j,\sigma} \hat{a}^{}_{j,\sigma}$, and is 
used to fix the particle number of the system.

The final term is the on-site interaction between electrons 
on the same site.  When two electrons, of opposite spin, occupy 
the same site the energy is raised or lowered by an amount $U$, 
due to forces such as the Coulomb repulsion and exchange forces.
The parameter $U$ thus gives the interaction strength, and can 
be repulsive $U>0$, or attractive $U<0$. 

Transforming the Hubbard Hamiltonian into a momentum 
representation using plane wave modes,
\begin{equation}
\hat{H}_0 + \hat{H}_{int}  = \sum_{k,\sigma} (\mu -2 \tau \cos ka ) 
\; \hat{c}^{\dag}_{k,\sigma} \hat{c}^{}_{k,\sigma} + 
U \sum_{k,k',q} \hat{c}_{k+q,\uparrow}^{\dag} \hat{c}^{}_{k,\uparrow}
\hat{c}_{k'-q,\downarrow}^{\dag} \hat{c}^{}_{k',\downarrow} \; , 
\end{equation}
we obtain the dispersion relation $\epsilon_k = -2\tau \cos ka$, 
where the hopping matrix element, $\tau$, determines the 
bandwidth $\epsilon_{max} - \epsilon_{min} = 4\tau$.

\subsection{Bosonization of the Hubbard Model}

Currently, bosonization of the Hubbard model relies on the 
projection of the model onto the Luttinger model. The 
transformation from a lattice model to the continuum may be 
achieved by a number of methods (Emery 1979, Shankar 1995,
Schulz 1993). The Hamiltonian splits into separate bosonic 
Hamiltonians for the charge and spin sectors.

Away from half filling the charge sector of the model becomes 
a Luttinger model.  The spin sector, on the other hand, contains 
a Luther-Emery, backward scattering, interaction term.  For a 
repulsive on-site interaction, $U>0$, this term is irrelevant 
under renormalization, and does not affect the large distance 
asymptotic form of the correlations.  In the case of an 
attractive on-site interaction, $U<0$, the interaction term 
remains relevant, and a spin gap is present.

At half filling, the charge sector also gains a Luther-Emery term. 
In this case the opposite applies to the spin sector, a repulsive 
term, $U>0$, generates a charge gap, whereas the attractive case, 
$U<0$, is irrelevant under renormalization.  The generation of the 
charge gap defines the transition as a Mott type ``metal-insulator'' 
transition (Gul\'acsi and Bedell 1994a). 

\subsubsection{Linearisation of the Hamiltonian}

As the methodology of bosonization developed previously does 
not rely on a continuum model, in a first approximation it is 
possible to directly bosonize the model by linearising the 
dispersion.  This gives a Tomonaga model,
\begin{equation}
\hat{H} \simeq v_F \sum_{k,\sigma} (|k|-k_F) \; 
\hat{c}^{\dag}_{k,\sigma} \hat{c}^{}_{k,\sigma} 
+ U \sum_{q} \sum_{k,k'} \hat{c}_{k+q,\uparrow}^{\dag}
\hat{c}^{}_{k,\uparrow} \hat{c}_{k'-q,\downarrow}^{\dag}
\hat{c}^{}_{k',\downarrow}
\end{equation}
with $v_F =2\tau \sin k_Fa$.  This may then have positron 
states appended to give a Luttinger type kinetic term. 

The interaction terms away from half-filling are easily 
obtainable for small positive $U$, where the interactions 
may be treated as a perturbation.  We then have,
\begin{eqnarray}
& g_{2\varrho} = \frac{U}{2\pi} 
\; , \; \; \; g_{4\varrho} = \frac{U}{2\pi} 
\nonumber \\
& g_{2\chi} = -\frac{U}{2\pi} 
\; , \; \; \; g_{4\chi} = -\frac{U}{2\pi} \; , 
\end{eqnarray}
which gives the relations for the exponents,
\begin{equation}
K_{\varrho} = \left( 1+\frac{U}{\pi v_F} \right)^{- 1/2} 
\; , \; \; \; 
K_{\chi} = \left( 1-\frac{U}{\pi v_F} \right)^{- 1/2} \; . 
\end{equation}
For small values of $U/\tau$, these approximations are valid 
when compared to the exact results (Schulz 1993). For strong 
interactions, or large $U/\tau$, the perturbative approximation 
is no longer reliable, and the true values of the correlation 
exponents may be obtained from the Bethe Ansatz solution, or 
conformal field theory results.

\subsubsection{Physical Fermions}

As we are interested only in the asymptotic low energy properties 
of the system the physical fermions of the model are defined in a 
way that only includes the components near to the Fermi points.  
We therefore define the physical fermions as,
\begin{equation}
\hat{\Psi}_{\sigma}(x) \simeq \hat{\psi}_{\sigma L}(x) + 
\hat{\psi}_{\sigma R}(x) \; , 
\end{equation}
where the left and right moving fermions are presumed to have 
a finite momentum cutoff around the Fermi points.  That is,
\begin{equation}
\hat{\psi}_{\sigma L}(x) \simeq \sum_{k=-k_F-\Lambda}^{-k_F+\Lambda} 
e^{ikx}d_{k\sigma} \; , 
\end{equation}
and similarly for the right movers.

A representation of the physical fermions in terms of a 
non-local combination of left and right movers is expressed 
by Voit (1995), where,
\begin{equation}
\hat{\Psi}^{\dag}_{\sigma}(x) = -\frac{i}{2L} 
\int_{-L/2}^{L/2} \left( \hat{\psi}^{\dag}_{\sigma L}(x+y) + 
\hat{\psi}^{\dag}_{\sigma R}(x+y) \right) \cot \frac{\pi y}{L} 
\; dy \; . 
\label{fermion_rep}
\end{equation}
The solution of the resultant integrals giving the correlation 
functions obtained from the representation in 
Eq.\ (\ref{fermion_rep}) proves to be a difficult task, 
and the qualitative agreement of the simplified approximation 
with known limits and results does not necessitate their 
calculation, and hence, evaluation is considered a future task. 

\subsection{Correlation and Green's Functions}

In combining operators for the physical fermions we obtain 
four combinations of left and right movers,
\begin{equation}
\hat{\Psi}^{}_{\sigma}(x) \hat{\Psi}^{\dag}_{\sigma}(0) 
\simeq \hat{\psi}^{}_{\sigma L}(x) \hat{\psi}^{\dag}_{\sigma L}(0) 
+ \hat{\psi}^{}_{\sigma R}(x) \hat{\psi}^{\dag}_{\sigma L}(0) 
+ \hat{\psi}^{}_{\sigma L}(x) \hat{\psi}^{\dag}_{\sigma R}(0) 
+ \hat{\psi}^{}_{\sigma R}(x) \hat{\psi}^{\dag}_{\sigma R}(0) \; .
\end{equation}
The terms containing mixtures of left and right moving fermions 
correspond to excitations with $q\simeq 2k_F$, whereas the terms 
with two of the same fermion type relate to the $q\simeq 0$ 
components.  We can therefore neglect the $2k_F$ terms as not 
relevant to the low energy physics.

The asymptotic form of the finite temperature correlation 
functions for the physical fermions as a sum of left and 
right moving terms, becomes,
\begin{eqnarray}
\langle \hat{\Psi}^{}_{\sigma}(x) \hat{\Psi}^{\dag}_{\sigma}(0) 
\rangle_{\beta} 
& \simeq \langle \hat{\psi}^{}_{\sigma L}(x) 
\hat{\psi}^{\dag}_{\sigma L}(0) + \hat{\psi}^{}_{\sigma R}(x) 
\hat{\psi}^{\dag}_{\sigma R}(0) \rangle_{\beta} 
\nonumber \\
&= e^{i k_Fx} \prod_{\mu = \varrho , \chi} 
\frac{\pi}{\sqrt{\beta v_{\mu} \sinh \frac{\pi x}{\beta v_{\mu}}}} 
\left( \frac{\pi}{\beta v_{\mu} \sinh \frac{\pi x}{\beta v_{\mu}}} 
\right)^{\gamma_{\mu}} \; . 
\end{eqnarray}

The correlation functions allow us to calculate the single particle 
Green's function for the Hubbard model, from,
\begin{eqnarray}
& G_{\sigma \sigma}^{\beta} (x,t) = -i \theta(t) 
\left\langle \left\{ \hat{\Psi}^{}_{\sigma}(x,t), 
\hat{\Psi}^{\dag}_{\sigma}(0) \right\} \right\rangle_{\beta} 
\nonumber \\
&= -i \theta(t) \left( \langle \hat{\Psi}^{}_{\sigma}(x,t)
\hat{\Psi}^{\dag}_{\sigma}(0) \rangle_{\beta} + 
\langle \hat{\Psi}^{}_{\sigma}(-x,-t) 
\hat{\Psi}^{\dag}_{\sigma}(0) \rangle_{\beta} \right) \; , 
\end{eqnarray}
we obtain,
\begin{eqnarray}
& G_{\sigma \sigma}^{\beta} (x,t) \simeq 
-i \theta (t) \cos k_F x \Bigg[ \prod_{\mu = \varrho , \chi} 
\left( \frac{\pi}{\beta v_{\mu} \sinh 
\frac{\pi (x+v_{\mu}t)}{\beta v_{\mu}}} 
\right)^{\frac{\gamma_{\mu}}{2}} 
\left( \frac{\pi}{\beta v_{\mu} \sinh 
\frac{\pi (x-v_{\mu}t)}{\beta v_{\mu}}} 
\right)^{\frac{1+\gamma_{\mu}}{2}} 
\nonumber \\
& ~~~ \nonumber \\
& + \prod_{\mu = \varrho , \chi} 
\left( \frac{\pi}{\beta v_{\mu} \sinh 
\frac{\pi (x-v_{\mu}t)}{\beta v_{\mu}}} 
\right)^{\frac{\gamma_{\mu}}{2}} 
\left( \frac{\pi}{\beta v_{\mu} \sinh 
\frac{\pi (x+v_{\mu}t)}{\beta v_{\mu}}} 
\right)^{\frac{1+\gamma_{\mu}}{2}} \Bigg] \; , 
\end{eqnarray}
where each term in the sum is in qualitative agreement with 
the results from conformal field theory (Frahm and Korepin 1990).

In taking the zero temperature limit for the equal time Green's 
function we recover the form,
\begin{equation}
G_{\sigma \sigma}(x) \simeq x^{- 1 - \alpha} \; , 
\end{equation}
in this case for $\alpha = 1+\gamma_{\varrho} + \gamma_{\chi}$, 
which is in agreement with the known results (Kawakami and
Yang 1991, Christe and Henkel 1993, Gul\'acsi 1997). 

\subsection{Bosonization and Conformal Field Theory}

The results obtained from conformal field theory are expressed 
as the correlations of the conformal fields themselves, and the 
physical fields cannot easily be related to these.  However, the 
exponents of the leading terms of the asymptotic expansion of 
the physical fields may be approximated by the conformal method.

The almost identical relation between the correlations of the 
fermion operators in our bosonization approach, and the correlations 
of the conformal fields may appear to be due to the simple relation 
we have used between the physical fermions and the constructed left 
and right moving fermion fields.  The left and right moving fields 
are independent and conformally invariant, each with a separate 
fundamental velocity.  The relationship between the physical 
fermions and left and right moving terms may therefore give 
insight into the relationship of conformal and physical fields 
in the conformal approach.

The agreement between the long distance form in the bosonization 
and conformal approaches shows that the bosonization approach 
is appropriate for obtaining the asymptotic correlations.  
Models that contain gaps in their excitation spectrum are no 
longer conformally invariant, and therefore cannot be solved 
using the methods of conformal field theory.  The extension of 
the bosonization method to systems with gapped excitations, such 
as the Luther-Emery model, may therefore give the asymptotic 
form of the correlation functions to a good approximation, 
where they cannot be obtained from the conformal approach.  
This avenue is currently being explored.

\subsection{The Metal-Insulator Transition at Finite Temperature}

The metal-insulator transition has been examined at zero 
temperature (Gul\'acsi and Bedell 1994a), 
using bosonization of a two band model.  The lower band is 
the Luttinger band, whilst the upper band contains the 
Luther-Emery term.  A zero temperature single particle Green's 
function has been proposed for the Hubbard model, by the same 
authors, again using the two band bosonization method 
(Gul\'acsi and Bedell 1994b). 

Given a result for the finite temperature Green's function, 
it may be compared to the zero temperature case proposed by 
Gul\'acsi and Bedell (1994a).  The temperature dependence 
of the Mott transition in the Hubbard model may then be examined.
From the exact Luther-Emery model, and solutions in various 
limits (Voit 1998) has proposed an ansatz form for 
the single particle spectral density function of models in 
this class.  From this proposed function it may be possible 
to calculate the corresponding finite temperature Green's 
functions to be used as an guide.

Using our bosonization method we have so far been unable to 
calculate the Green's functions of the Luther-Emery model, 
although the work is still in progress.  This is due to the 
complexity of the representation of the spinless fermion 
operators in terms of the physical fermions or their boson 
representation.

\section{Conclusions}

The Luttinger liquid characteristics 
describe the ground state properties of interacting one
dimensional systems. All the experimental data, however, 
are finite temperature measurements. Because of this it 
has become timely to comprehensively describe the finite 
temperature characteristics of Luttinger liquids. As 
a first step in achieving this goal, we extended the 
well-known bosonization method of one dimensional 
electron systems to finite temperatures. 

We have reviewed the derivation of the bosonization 
operator identity in one dimension.  The formalism was then used 
to examine the properties of a simple linear-dispersion fermion 
model, both with and without spin.  The existence of spin-charge 
separation was shown for the interacting case, showing that the 
interactions between fermions at the same point of the Fermi 
surface is the mechanism behind this phenomenon.

The single particle correlation functions, from which Green's 
functions may be derived, were calculated explicitly to avoid 
problems of normal ordering of the operators in exponentials.  
Previous calculations have relied on the definition of non-normal 
ordered boson fields, which required a convergence factor to allow 
solution of the resultant integrals. In using the normal ordered 
representation of von Delft and Sch\"{o}ller (1998), and 
the iterative solution of expectations given by Mermin (1966),
we were able to explicitly retain apparent divergences and 
recombine these terms to obtain our results.

The solution to the Tomonaga-Luttinger model was examined, 
to show the generation of anomalous dimensions of operators 
which signal the non-universal power law behaviour in correlation 
functions.  The dependence of this phenomenon was shown to rely 
on the scattering of fermions of opposite spin on separate branches 
on the Luttinger model. Correlation functions were again explicitly 
derived, where a mixing of the correlations between separate fermion 
species is observed.

The asymptotic form of the finite temperature correlation and 
Green's functions for the Hubbard model were calculated.  As 
far as we are aware, these have not previously been derived 
using the bosonization method, and our results agree with the 
form obtained from conformal field theory. It appears, 
however, that the nature of the relationship between the 
physical fermions of the theory and the constructed boson 
and fermion fields requires further examination.

In determining the finite temperature behaviour of a two 
band Hubbard model we hoped to model the transition between 
a Luttinger liquid behaviour at high temperature and a 
Luther-Emery liquid at low temperature.  At high temperatures 
it is assumed that the thermal excitations will make the gap 
in the spectrum of the excitations irrelevant, and the model 
will exhibit Luttinger liquid behaviour, but at lower 
temperature the gap becomes relevant and the behaviour 
undergoes a phase change to a Luther-Emery liquid.

Whether the high temperature behaviour of a gapped degree 
of freedom indeed behaves as a Luttinger liquid with perhaps 
renormalized parameters is the desired goal of this line of 
enquiry. The generalization of finite temperature behaviour 
of the Luther-Emery model to other models within the class 
is presumed from the renormalization calculations showing 
agreement for this class of models at zero temperature.

Difficulties appear with the lack of a simple representation 
of the original fermion operators, in terms of the new 
spinless fermions, as well as the restriction of 
refermionization to bosonized Hamiltonians with a 
linear spectrum.

Further examination of the correlation functions for the 
Luther-Emery model may show whether the iterative solution 
is tractable in solving the single particle correlator.  
This would allow a simple substitution into the two band 
model for the Mott metal-insulator transition.  Due to the 
equivalence of models, this would also give the single 
particle finite temperature correlation function for the 
sine-Gordon and massive Thirring models, and hence the 
importance of such solution may be seen.

The Hubbard model is soluble at finite temperature by use 
of the Thermodynamic Bethe Ansatz.  A comparison of 
the finite temperature properties of the Hubbard model, 
and those of the Tomonaga-Luttinger model, and possibly 
the Luther-Emery model, may be used to clarify the Luttinger 
liquid conjecture for systems at finite temperature.

\newpage

\section*{References}

~ \\
Abrikosov, A. A., Gorkov, L. P., and Dzyaloshinski, L. P., 1963,   
{\it{ Methods of quantum field theory in statistical physics}}, 
Prentice-Hall. \\
Andrei, N. and Lowenstein, J. H., 1982, Phys. Rev. Lett. 
{\bf 46}, 356. \\
Andrei, N., Furuya, K. and Lowenstein, J. H., 1983, Rev. Mod. Phys. 
{\bf 55}, 331. \\
Bowen, G., Gul\'acsi, M. and Rosengren, A. 2000, Aust. J. Phys.
{\bf 53}, 553. \\
Cardy, J. L., 1984, J. Phys. A{\bf 17}, L385. \\
Christe, P., and Henkel, M., 1993, {\it{ Introduction to Conformal 
Invariance and Its Applications to Critical Phenomena}}, 
Springer-Verlag. \\
Cloiseaux, J. des, and Pearson, J., 1962, Phys. Rev. 
{\bf 128}, 2131. \\
Delft, J. von, and Sch{\"o}ller, H., 1998, Ann. Phys. 
{\bf 4}, 225. \\
Dover, C. B., 1968, Ann. Phys. {\bf 50}, 500. \\
Efetov, K. B. and Larkin, A. I., 1975, Zh. Eksp. Teor. Fiz.
{\bf 69}, 764 [Sov. Phys. - JETP {\bf 42}, 390 (1976)]. \\
Emery, V. J., 1979, in {\it{ Highly Conducting One-Dimensional 
Solids}}, ed. J. T. Devreese {\sl et al.}, Plenum Press. \\
Filyov, V. M., Tsvelick, A. M. and Wiegmann, P. B., 1981, 
Phys. Lett. A{\bf 81}, 175. \\
Filyov, V. M., Tsvelick, A. M. and Wiegmann, P. B., 1982, 
Phys. Lett. A{\bf 89}, 157. \\
Frahm, H., and Korepin, V. E., 1990, Phys. Rev. B{\bf 42}, 10553. \\
Garrod, C., 1995, {\it{ Statistical Mechanics and Thermodynamics}},
Oxford University Press. \\
Gaudin, M, 1971, Phys. Rev. Lett. {\bf 26}, 1301. \\
Griffiths, R. B., 1964, Phys. Rev. {\bf 133}, A768. \\
Gul\'acsi, M., 1997, Phil. Mag. B{\bf 76}, 731. \\
Gul\'acsi, M., and Bedell, K. S., 1994a, Phys. Rev. Lett. {\bf 72}, 
2765. \\
Gul\'acsi, M., and Bedell, K. S., 1994b, Physica B{\bf 199}, 492. \\
Gutzwiller, M., 1963, Phys. Rev. Lett. {\bf 10}, 159. \\
Haldane, F. D. M., 1979, J. Phys. C{\bf 12}, 4791. \\
Haldane, F. D. M., 1981, J. Phys. C{\bf 14}, 2585. \\
Hubbard, J., 1963, Proc. Roy. Soc. (London) {\bf A 276}, 238. \\
Inkson, J. C., 1984, {\it{ Many-body theory of solids}}, 
Plenum Press. \\
Katsura, S., 1965, Ann. Phys. {\bf 31}, 325 (1965).  \\
Kawakami, N. and Okiji, A., 1982, Solid State Commun. 
{\bf 43}, 467. \\
Kawakami, N., and Yang, S. -K., 1991, J. Phys. Cond. Matt. {\bf 3}, 
5983. \\
Korepin, V. E., Bogoliubov, N. M., and Izergin, A. G., 1993, 
{\it{ Quantum Inverse Scattering method and Correlation
Functions}}, Cambridge University Press. \\
Kr{\"o}nig, R. de L., 1935, Physica {\bf 2}, 968. \\
Lai, C. K., 1971, Phys. Rev. Lett. {\bf 26}, 1472. \\
Lai, C. K., 1973, Phys. Rev A{\bf 8}, 2567 (1973). \\ 
Lieb, E. H., and Wu, F. Y., 1968, Phys. Rev. Lett. {\bf 20}, 1445. \\
Luther, A., and Peschel, I., 1974, Phys. Rev. B{\bf 9}, 2911. \\
Luttinger, J. M., 1960, Phys. Rev. {\bf 119}, 1153. \\
Luttinger, J. M., 1963, J. Math. Phys. {\bf 4}, 1154 (1963). \\
Mahan, G. D., 1981, {\it{ Many-Particle Physics}}, Plenum Press. \\
Mattis, D. C. and Lieb, E. H., 1965, J. Math. Phys. {\bf 6}, 304. \\
Meden, V., and Sch{\"o}nhammer, K., 1992, Phys. Rev. B{\bf 46}, 
15753. \\ 
Mermin, N. D., 1966, J. Math. Phys. {\bf 7}, 1038. \\
Metzner, W., and Di Castro, C., 1993, Phys. Rev. B {\bf 47}, 16107. \\
Okiji, A. and Kawakami, N., 1983, Phys. Rev. Lett. {\bf 50}, 1157. \\
Ovchinnikov, A. A., 1969, Zh. Eksp. Teor. Fiz. {\bf 56}, 1354 
[Sov. Phys. - JETP {\bf 29}, 727 (1969)]. \\
Overhauser, A. W., 1965, Physics {\bf 1}, 307. \\
Rasul, J. W., 1982,  in {\it Valence Instabilities}, eds. P. Wachter
and H. Boppart, North Holland, Amsterdam, p. 49. \\
Ryzhik, I. M., and Gradshteyn, I. S., 1965, {\it{ Table of 
Integrals, Series, and Products}}, Academic Press. \\
Schlottmann, P., 1984, Z. Phys. B{\bf 55}, 293 (1984). \\
Schlottmann, P., 1993, J. Phys. Cond. Matter {\bf 5}, 5869. \\
Sch{\"o}nhammer, K. and Meden, V., 1996, Am. J. Phys. {\bf 64}, 
1168. \\
Schulz, H. J., 1993, in {\it{ Correlated Electron Systems}},
ed. V. J. Emery, World Scientific. \\
S{\'e}n{\'e}chal, D., 1999, cond-mat/9908262. \\
Shankar, R., 1995, Act. Phys. Pol. B {\bf 26}, 1835. \\
Spain, B., and Smith, M. G., 1970, {\it{ Functions of Mathematical 
Physics}}, Van Nostrand Reinhold Company. \\
Takahashi, M., 1971a, Prog. Theor. Phys. {\bf 46}, 401. \\
Takahashi, M., 1971b, Prog. Theor. Phys. {\bf 46}, 1388. \\
Takahashi, M., 1972, Prog. Theor. Phys. {\bf 47}, 69. \\
Takahashi, M. and Suzuki, M., 1972, Prog. Theor. Phys. 
{\bf 48}, 2187. \\
Theumann, A., 1967, J. Math. Phys. {\bf 8}, 2460. \\
Tomonaga, S., 1950, Prog. Theor. Phys. {\bf 5}, 544. \\
Tsvelick, A. M. and Wiegmann, P. B., 1982a, J. Phys. C{15}, 1707. \\
Tsvelick, A. M. and Wiegmann, P. B., 1982b, Phys. Lett. 
A{\bf 89}, 368. \\
Tsvelick, A. M. and Wiegmann, P. B., 1983a, Adv. Phys. 
{\bf 32}, 453. \\
Tsvelik, A. M. and Wiegmann, P. B., 1983b, J. Phys. 
C{\bf 16}, 2281. \\
Tsvelik, A. M. and Wiegmann, P. B., 1983c, J. Phys. 
C{\bf 16}, 2321. \\ 
Voit, J., 1993, Phys. Rev. B{\bf 47}, 6740. \\
Voit, J., 1995, Rep. Prog. Phys. {\bf 58}, 977. \\
Voit, J., 1998, Euro. Phys. J. B{\bf 5}, 505. \\
Witten, E., 1984, Commun. Math. Phys. {\bf 92}, 455. \\
Yang, C. N. and Yang, C. P., 1969, J. Math. Phys. 
{\bf 10}, 1115. \\
Yang, C. P., 1970, Phys. Rev. A{\bf 2}, 154. \\

\end{document}